\documentclass[12pt]{emulateapj}
\usepackage[]{natbib}
\usepackage{graphicx, amsmath}
\def\etal{{et~al.\null}}
\def\eg{{e.g.,}}
\begin{document}

\title{{\it 3D-HST} Emission Line Galaxies  at $z \sim 2$: Discrepancies in the Optical/UV Star Formation Rates}
\author{Gregory R. Zeimann \altaffilmark{1,2}, Robin Ciardullo \altaffilmark{1,2}, Henry Gebhardt \altaffilmark{1,2}, Caryl Gronwall \altaffilmark{1,2},
Donald P. Schneider \altaffilmark{1,2}, Alex Hagen \altaffilmark{1,2}, Joanna S. Bridge \altaffilmark{1,2}, John Feldmeier \altaffilmark{3}, and Jonathan R. Trump \altaffilmark{1,2,4}}
\altaffiltext{1}{Department of Astronomy \& Astrophysics, The Pennsylvania State University, University Park, PA 16802}
\altaffiltext{2}{Institute for Gravitation and the Cosmos, The Pennsylvania State University, University Park, PA 16802}
\altaffiltext{3}{Department of Physics and Astronomy, Youngstown State University, Youngstown, OH 44555}
\altaffiltext{4}{Hubble Fellow}

\begin{abstract}
We use {\sl Hubble Space Telescope\/} near-IR grism spectroscopy to examine the H$\beta$ line strengths of 
260 star-forming galaxies in the redshift range $1.90 < z < 2.35$.   We show that at these 
epochs, the H$\beta$ star formation rate (SFR) is a factor of $\sim$ 1.8 higher than what would be expected from the
systems' rest-frame UV flux density, suggesting a shift in the standard conversion between these quantities
and star formation rate.  We demonstrate that at least part of this shift can be attributed to metallicity, as
H$\beta$ is more greatly enhanced in systems with lower oxygen abundance.  This offset must be 
considered when measuring the star formation rate history of the universe.   We also show that
the relation between stellar and nebular extinction in our $z \sim 2$ sample is consistent with that observed in the local universe.

\end{abstract}
\label{sec:abs}

\section{Introduction}
\label{sec:intro}

Star formation and the build-up of stellar mass are key astrophysical parameters for our understanding 
of galaxy evolution and formation.  By quantifying the amount of star formation over a given time scale and 
co-moving volume, we can constrain models of galaxy formation, chemical enrichment (both interior and exterior
to galaxies), and the ionization history of the intergalactic medium \citep[\eg][]{tinsley80, madau96}. 
However, the process of converting observables, such as a galaxy's Balmer emission or UV flux density, into
actual star formation rates (SFRs) is fraught with difficulty, since virtually all common SFR indicators are 
indirect and sensitive only to the presence of high-mass, short-lived stars.  Assumptions concerning the shape 
of the initial mass function (IMF), the star formation history of the population, and the metal abundance of
the stars all play a role in translating measurable quantities into physical star formation rates.       

\citet{kennicutt98} and \citet{kennicutt+12} have reviewed the commonly used SFR indicators,
their calibration, and their limitations.  Two of the most useful of these are a galaxy's UV luminosity
density and the energy emitted in its Balmer lines.  The former is produced by stars with 
$M \gtrsim 5 M_{\odot}$, and therefore records star formation over the last $\sim 100$~Myr.   The 
method is extremely sensitive to dust, as just a few tenths of differential reddening can lead to magnitudes 
of extinction, and its calibration depends on a number of assumptions, including that of 
the population's IMF, metallicity, and star formation rate history.   Nevertheless, it is the technique
most commonly used for SFR measurements in the high-redshift universe.  In contrast, emission lines 
such as H$\alpha$ and H$\beta$ are the result of the ionizing photons produced by stars with 
$M \gtrsim 15 M_{\odot}$, and thus probe only the most recent episode of star formation, i.e., stars with
ages of $t < 10$~Myr.  Like the UV, Balmer emission also depends on the population's IMF, metallicity, 
and extinction, but the reaction of these lines to changes in the stellar population parameters is different.  As a result a
comparison of the two indicators can provide insights into both the stellar population and extinction,
even in the absence of additional information \citep[\eg][]{lee+09, meurer09}.

In this paper, we use near-IR spectroscopy with the {\sl Hubble Space Telescope (HST)\/} and a wealth of 
ancillary photometric data to examine the H$\beta$ line strengths and rest-frame UV flux densities of 260
star-forming galaxies in the redshift range $1.90 < z < 2.35$.  In \S2, we discuss the data for a combined $\sim$350 arcmin$^2$ region of the 
GOODS-N, GOODS-S, and COSMOS fields, and the reduction techniques needed to measure total H$\beta$ fluxes for galaxies with star formation
rates as low as $\sim 2 \, M_{\odot}$~yr$^{-1}$.  In \S3, we describe the procedures used to identify a
complete, H$\beta$-selected sample of objects at $z \sim 2$, and present our measurements of these
galaxies.  In \S 4, we calculate the galaxies' SFRs using the conversion factors summarized in \citet{kennicutt+12} 
and the local starburst galaxy extinction law found by \citet{calzetti01}.  We show that there is an 
inconsistency between the two SFRs, and examine the various parameters which might 
explain the offset.   We demonstrate that galactic metallicity is in large part responsible for the SFR 
discrepancy, and that a \citet{calzetti01} law reproduces the ratio of 
nebular-to-stellar extinction.  We conclude by discussing the possible impact of these measurements on 
other investigations of the high-redshift universe. For this work we adopt a standard $\Lambda$CMD cosmology, 
with $\Omega_{M} = 0.3$, $\Omega_{\Lambda} = 0.7$, and $H_0 = 70$~km~s$^{-1}$~Mpc$^{-1}$.

\section{Data and Reductions}
\label{sec:obs}

Our study of star formation in the $z \sim 2$ universe is focused on three
$\sim 120$~arcmin$^2$ patches of sky in the COSMOS \citep{COSMOS}, GOODS-N,
and GOODS-S \citep{GOODS} fields.  In these regions, there is a wealth of
photometry and spectroscopic data available for analysis, including broadband photometry from a
host of space missions, broad- and intermediate-band photometry from the ground, and
optical and near-IR slitless spectroscopy from {\sl HST.}  Below
we describe the data used in our analysis.

\subsection{Optical/Near-IR Imaging}

To perform our analysis, we took advantage of the \citet{skelton2014} photometric catalogs
produced by the 3D-HST project \citep[][GO-11600, 12177, and 12328]{3DHST}.
\citet{skelton2014}  homogeneously combined
147 distinct ground-based and space-based imaging data sets covering the
wavelength range $0.3 \mu$m to $8.0 \mu$m in five well-observed legacy fields,
including GOODS-N, GOODS-S, and COSMOS\null.  In their analysis,
\citet{skelton2014} obtained and reduced {\sl HST/}WFC3 images from both the
CANDELS \citep{CANDELS} and 3D-HST \citep{3DHST} surveys,
and created a source catalog using \textsc{SExtractor} \citep{bertin96} on
co-added F125W+F140W+F160W images.  These catalogs, detection segmentation
maps, point spread functions (PSFs), and flux enclosed in PSF-matched apertures were
then used to measure total flux densities in a wide variety of publicly available
imaging datasets. For $z \sim 2$ systems, these data offer unprecedented access to
the rest-frame ultraviolet and allow precision measurements of the slope of each object's
rest-frame UV continuum.

\subsection{{\it HST} Spectroscopy}

Our rest-frame optical emission-line measurements come from 3D-HST, a near-IR grism survey with the {\sl HST\/} WFC3 camera.  
The 3D-HST primary observations with the G141 grism consists of $R \sim 130$ slitless spectroscopy between 
$1.08~\mu{\rm m} < \lambda < 1.68~\mu$m over a 625~arcmin$^2$ region of sky, which includes 
$\sim 80\%$ of the CANDELS footprint; when combined with accompanying direct images through 
the F140W filter of WFC3, these data provide full coverage of the rest-frame wavelengths 3700-5020~\AA\ 
for all $1.90 < z < 2.35$ galaxies with unobscured emission line fluxes brighter than $10^{-17}$ ergs cm$^{-2}$ s$^{-1}$ (at 1-$\sigma$) which corresponds
to unobscured SFRs greater than 
$\sim 2 M_{\odot}$~yr$^{-1}$.  Included in this wavelength range are the strong emission lines of 
[O~II] $\lambda 3727$, [O~III] $\lambda\lambda 4959,5007$, [Ne~III] $\lambda\lambda
3869,3960$, and hydrogen (H$\beta$, H$\gamma$ and H$\delta$).  

To reduce these data, we began with the pre-processed, calibrated ``FLT'' files in the {\sl HST\/} Data
Archive.  These are the products of the automated reduction process {\tt calwf3}\footnote[3]{http://www.stsci.edu/hst/wfc3/documents/handbooks/}, 
which uses the latest reference files to measure and subtract the bias, correct for non-linearity, flag 
saturated pixels, subtract the dark image, divide by the flat field, calculate the gain, and apply the flux 
conversion.  This process was identical for both the direct and the grism images, with the 
exception of the flat fielding step:  the grism data were flat fielded at a later stage using the 
{\tt aXe}\footnote[4]{http://axe-info.stsci.edu/} software and a master sky flat.  

Each grism observation was accompanied by a shallow ($\sim 200$~s) F140W exposure, which served to
define the position of each object's wavelength zeropoint and trace, and hence facilitate spectral calibrations and
extraction.  These images were combined using the standard procedures of {\tt MultiDrizzle} \citep{fruchter09}, 
and then co-added with the deeper CANDELS F125W and F160W frames to match the detection image used 
in the photometric catalogs of \citet{skelton2014}.   These data were processed by \textsc{SExtractor}
to produce a master catalog of all objects containing more than five pixels above a 3-$\sigma$ 
per pixel detection threshold and having a total AB magnitude \citep{oke83} brighter than 26.   The positions
of these sources were then transformed back to the coordinate system of the shallower F140W 
image to enable 2-D spectral extractions on the grism frames.  Positional uncertainties from this process were $\lesssim 0.5$ pixel in the F140W frame.
The grism data were reduced using version 2.3 of the program {\tt aXe} \citep{kummel09}, in a manner
similar to that described in the WFC3 Grism Cookbook\footnote[5]{http://www.stsci.edu/hst/wfc3/analysis/grism\_obs/cookbook.html}.   The task
{\tt AXEPREP} was used to subtract the master sky frame\footnote[6]{http://www.stsci.edu/hst/wfc3/analysis/grism\_obs/\\calibrations/wfc3\_g141.html} from each image;
such a step is critical for the extraction of the faintest targets.  We do note that the background of a 
grism image is variable over time and best fit using a full set of master sky images; however, as we are 
solely concerned with the detection and measurement of emission lines, large-scale variations in the
continuum (of the order of $\lesssim$5\% of the original background) are not a serious issue.

After subtracting the sky background, we began the process of extracting the 2-D spectrum of every object in the
master \textsc{SExtractor} catalog.  This was done using {\tt AXECORE}, which defines each source's
extraction geometry, flat fields the region containing the spectral information,
applies the wavelength calibration, and determines the contamination from overlapping spectra.    Each 
object was traced with a variable aperture based on its size on the direct image ($\pm 4$ times the
projected width of the source in the direction perpendicular to the spectral trace).  Objects present on
multiple frames were processed by {\tt DRZPREP} and {\tt AXEDRIZZLE}, which rejected the 
cosmic rays, drizzled the data to a common system \citep{fruchter09}, and co-added the images into 
one higher signal-to-noise ratio 2-D spectrum.  Finally, the optimal
extraction method discussed by \citet{kummel09} was employed to create a 1-D spectrum for each object that
includes flux density, error on the flux density, and a contamination fraction in units of flux density.

To conclude our extraction process, we created a webpage that combined the 2-D grism 
images with the 1-D extracted spectra in a visually effective format.   This step was performed with the program
{\tt aXe2web}\footnote[7]{http://axe.stsci.edu/axe/axe2web.html}, which was used to convert an input
catalog and the {\tt aXe} output files into a summary of the full reduction.  Each object was displayed 
on a separate row with its magnitude, ($x,y$) position, equatorial coordinates,
direct image cutout, grism image cutout, and its 1-D extracted spectrum in counts and flux.  This webpage 
format provides an easy and efficient way to view a summary of the reductions, maintain quality 
control, and select subsamples of objects for science purposes.

\begin{figure}[htp] 
\includegraphics[width=.45\textwidth]{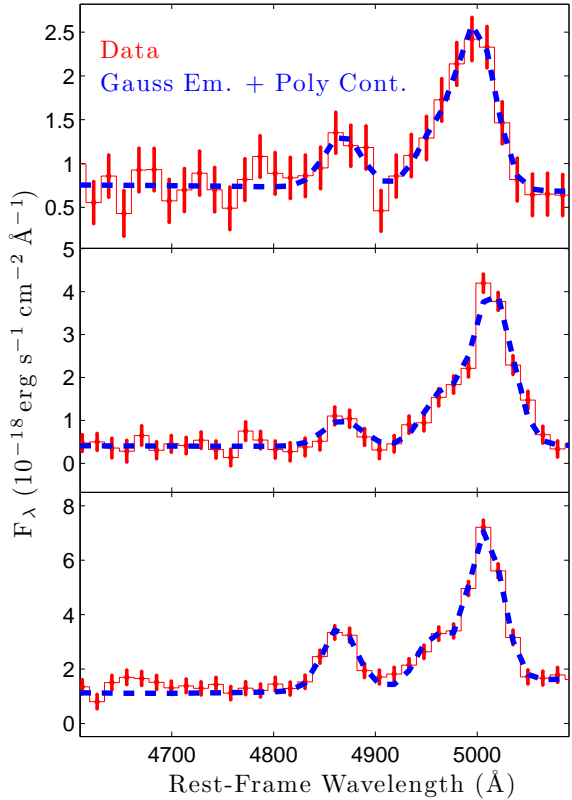}
\centering
\caption{Examples of the region about H$\beta$ and the [O~III] doublet for three typical $z \sim 2$ COSMOS field
galaxies observed with the WFC3 G141 grism of the 3D-HST program.  From top to bottom, the three spectra represent below average, average, and better than average
examples.  The emission lines and underlying
continuum were modeled with a Gaussian and polynomial, respectively.}
\label{fig:sample_spec}
\end{figure}

\vskip +1cm
\section{Sample Selection and Measurements}

We began our analysis by examining each 1-D extracted spectrum by eye to search for evidence of the
emission lines of hydrogen (H$\alpha$ and H$\beta$), [O~II] $\lambda 3727$, and [O~III]
$\lambda\lambda 4959,5007$.  (Other commonly detected lines included [Ne~III] $\lambda 3869$
and the [S~II] blend at $\lambda\lambda 6716,6731$.)  Spectra exhibiting two or more emission
lines were inspected in more detail.  Specifically, in the redshift range $1.90 < z < 2.35$, the bright
lines of [O~III], H$\beta$, [Ne~III], and [O~II] all fall within the coverage of the G141 grism.  Moreover,
the limited spectral resolution of the survey ($\sim 93$~\AA) blends the [O~III] $\lambda\lambda 4959, 5007$
doublet together, creating a distinctive asymmetric profile (see Figure~\ref{fig:sample_spec}).  As [O~III]
$\lambda 5007$ is typically the strongest emission line in these spectra, this was the most common 
feature selected for detailed inspection.  Secure redshifts were determined if two or more emission lines  provided a consistent redshift for the object.  (The [O~III] doublet counted as two lines, due to its unique shape.)  In total, we visually inspected $> 50,000$ spectra, and obtained redshifts for 323 $1.90 < z < 2.35$
galaxies with photometric coverage in the rest-frame UV. 

Overlapping spectra are a significant issue in slitless spectroscopy:  frequently, a portion of the 
dispersed order of one source will overlap the spectrum of another, causing ``contamination.''  
To model this effect, we used the sizes and magnitudes of every object in the \textsc{SExtractor} catalog
to create a 2-D Gaussian model of its expected spectrum \citep{kummel09}.  This map was then 
projected back onto the coordinate system of the science frame to create a contamination map
of the region.  Unfortunately, while this procedure is sufficient to identify most spectral 
superpositions, it does not identify or properly quantify all regions where the the systematics of contamination subtraction dominates the error in the continuum. In fact, a visual inspection of our sample of 323 $z \sim 2$ galaxies found 59 objects where the systematic error of contamination subtraction was greater than the statistical error of the target spectrum.  These objects were removed from our sample along with four galaxies that are likely active galactic nuclei 
(see \S 3.4), leaving a total of 260 $1.90 < z < 2.35$ galaxies distributed over the three fields of GOODS-S, GOODS-N, and COSMOS.

\begin{figure*}[htp] 
\includegraphics[width=1\textwidth]{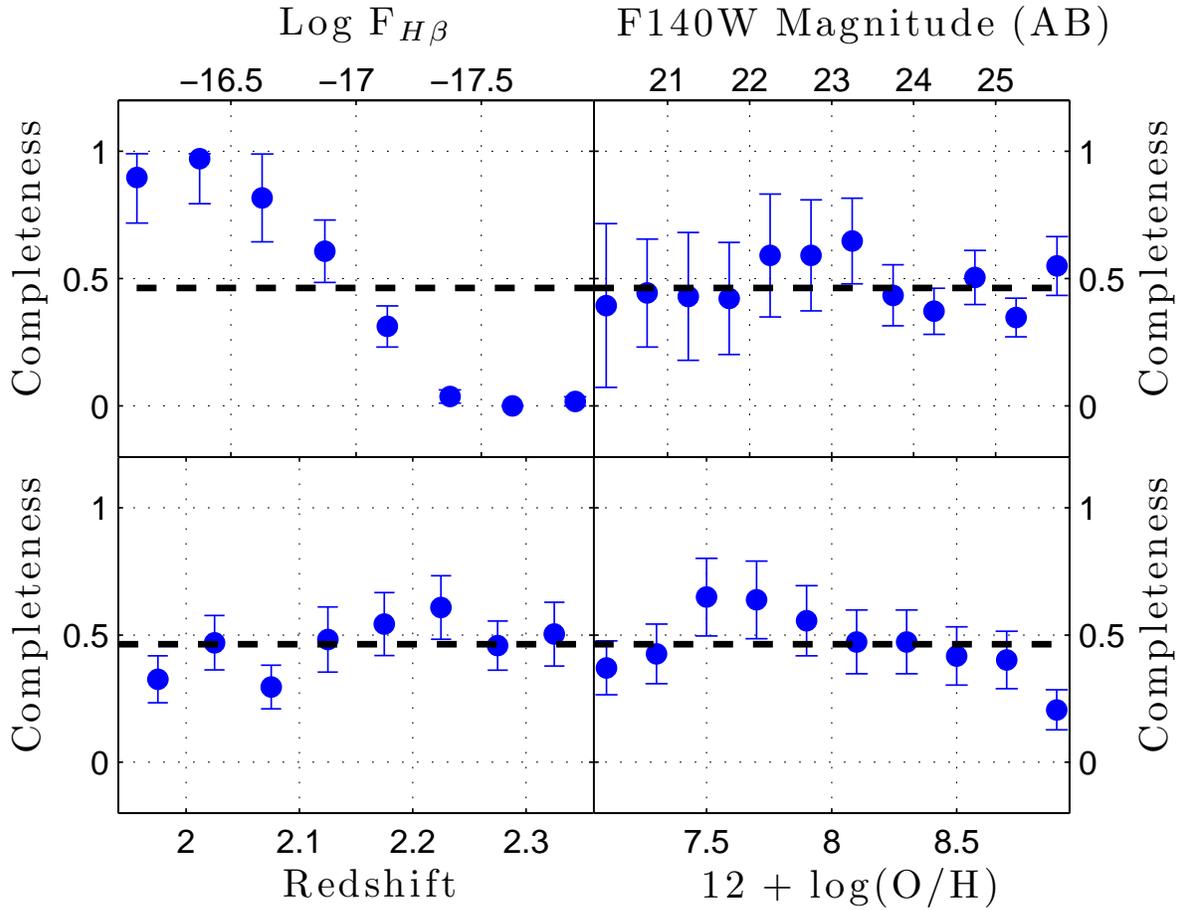}
\centering
\caption{Survey completeness and 1-$\sigma$ error bars as a function of H$\beta$ flux, continuum magnitude, redshift, 
and metallicity, as determined by a Monte Carlo experiment.  The dashed line shows 
the total recovery rate for our experiment.    Our recovery fraction is only a function of line flux, with the 
50\% completeness limit in the GOODS fields at $F_{{\rm H}\beta} \sim 10^{-17}$~ergs~s$^{-1}$~cm$^{-2}$.}
\label{fig:simulation}
\end{figure*}

To understand our sample selection, we ``observed'' a set of simulated emission-line spectra
in the exact same manner as our program data.  To realistically model uncertainties produced by contamination, we began with the  F140W magnitude and positional
distributions defined in the master \textsc{SExtractor} catalog.  We then randomly drew from a uniform distribution and assigned to each object a 
redshift ($1.90 < z < 2.35$), a metallicity ($7 < 12 + \log({\rm O/H}) < 9$), and H$\beta$ flux 
(drawn from a uniform distribution in log space with $-18 < \log F_{{\rm H}\beta} < -16$~ergs s$^{-1}$~cm$^{-2}$).  For a given metallicity and H$\beta$ flux, it is possible to use locally-calibrated, strong line metallicity indicators to predict the line strengths for [O~III], [Ne~III], and [O~II].  We used the polynomial relations in Table 4 and Equation 1 of \citet{maiolino+08}, which is discussed in more detail in \S 3.2, to convert metallicity and H$\beta$ flux into line strengths of [O~III], [Ne~III], and [O~II]. These lines were superposed onto a constant flux density continuum that matched the object's F140W magnitude.    
A total of 500 of these high-resolution template spectra were then placed onto a simulated grism image (and an accompanying 
direct image) using the {\tt aXeSim}\footnote[8]{http://axe.stsci.edu/axesim/} software package, and extracted 
in the same manner as the original data.  A summary of this analysis is shown in Figure~\ref{fig:simulation}.   
From the figure, it is clear that our ability to detect and measure H$\beta$ is virtually independent of 
redshift, metallicity, and continuum magnitude.  Formally, for the GOODS-N and GOODS-S fields, our  50\% and 80\% completeness limits for H$\beta$
are $\sim 10^{-17}$~ergs~s$^{-1}$~cm$^{-2}$ and $\sim 3 \times 10^{-17}$~ergs s$^{-1}$~cm$^{-2}$, respectively, with
little variation across parameter space.  Due to the higher background, the COSMOS limits are shallower by a factor of $\sim$ 1.5 \citep{3DHST}. Note that these 50\% limits are roughly equivalent to the 1-$\sigma$ 
flux measurement found by \citet{3DHST}.  Our high recovery fraction is due principally to the fact that
most of our galaxies were originally identified via the presence of much stronger emission lines, such 
as the [O~III] doublet and [O~II] $\lambda 3727$.  This allows for the identification and measurement of  
H$\beta$ to much lower flux limits than would be possible based on blind detection of H$\beta$.

\begin{figure*}[htp] 
\includegraphics[width=.48\textwidth]{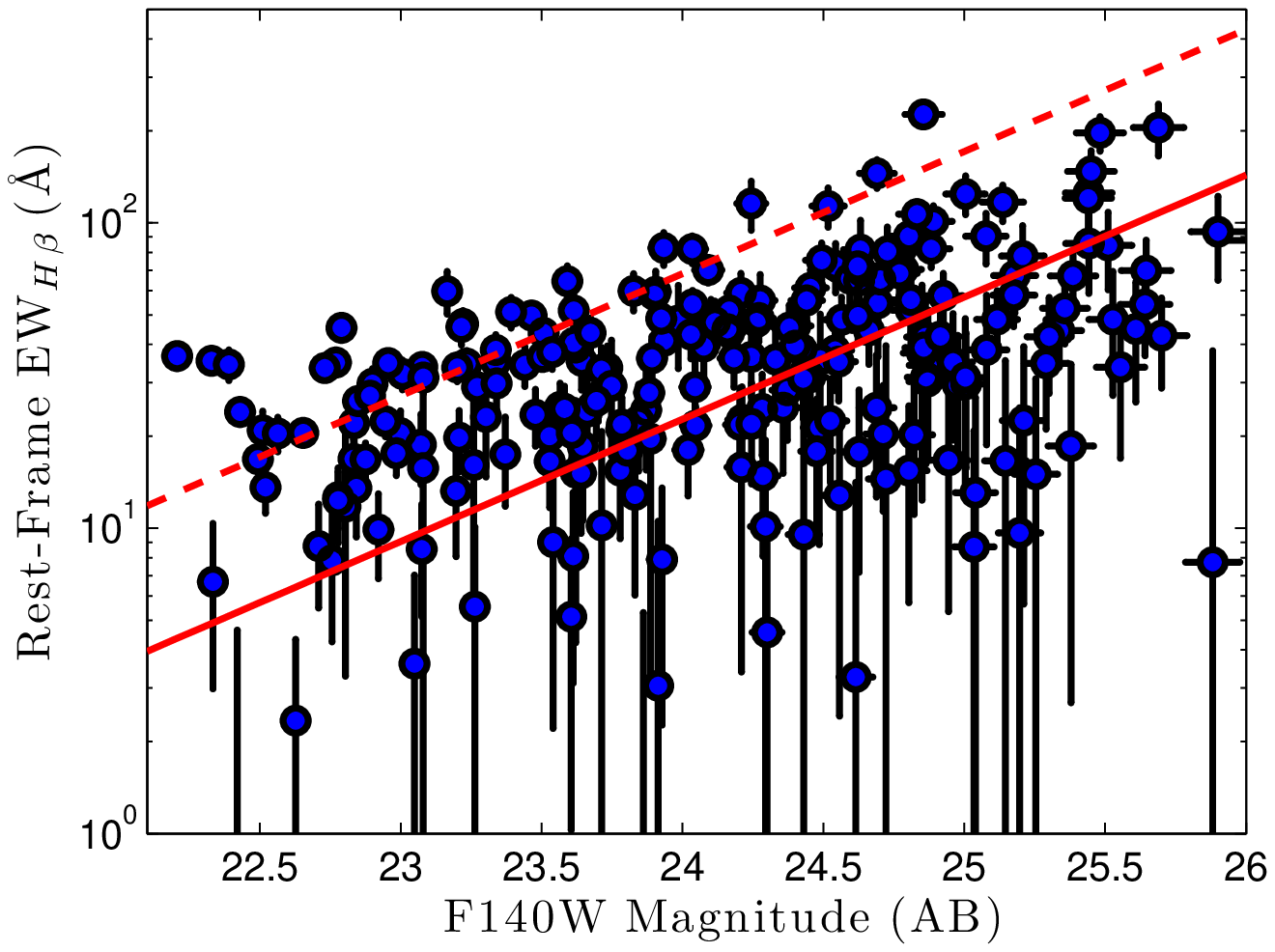}
\includegraphics[width=.48\textwidth]{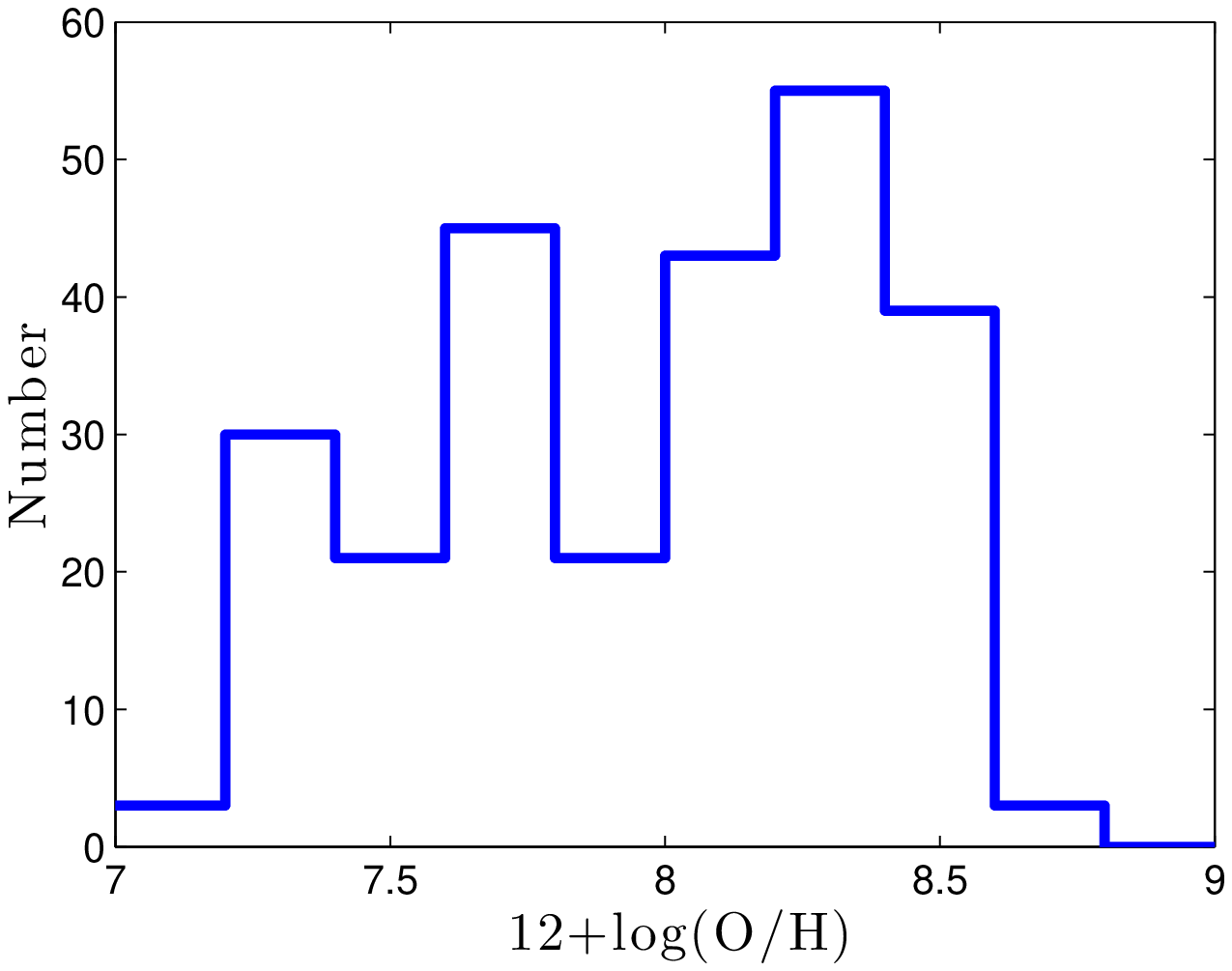}
\centering
\caption{{\it Left:} The rest-frame H$\beta$ equivalent widths (\AA) plotted against continuum AB magnitude for our
sample of $z \sim 2$ sources.  The solid red line is our 50\% completeness limit 
$F_{{\rm H}\beta} \sim 10^{-17}$~ergs~s$^{-1}$~cm$^{-2}$, while the red dashed line 
shows our 80\% completeness limit, $F_{{\rm H}\beta} \sim 3 \times 10^{-17}$~ergs~s$^{-1}$~cm$^{-2}$.  All equivalent widths were calculated using 
the measured flux of H$\beta$, the continuum magnitude measured on the direct F140W frame, and the
redshift.  We include 1-$\sigma$ error bars for each measurement.{\it Right:} Distribution of the best-fit gas-phase metallicities ($12+ \log{\rm (O/H)}$)
from \citet{gebhardt+14}.}
\label{fig:ew_mag}
\end{figure*}

\vskip +1cm
\subsection{H$\beta$ Luminosity}

The H$\beta$ line luminosities were determined by fitting the continuum of each $z \sim 2$ spectrum with a 
fourth-order polynomial, while simultaneously fitting Gaussians of a common width and redshift to the emission
lines of [O~II] $\lambda 3727$, [Ne~III] $\lambda 3869$, H$\gamma$, H$\beta$, [O~III] $\lambda 4959$, and [O~III] $\lambda 5007$.  The fourth-order polynomial was used due to the possible presence of a 4000 \AA\ break. However, we also fit first-order polynomials and Gaussians in small wavelength windows for [O~II], [NeIII], H$\gamma$, and the combination of H$\beta$ and [O~III] due to their proximity.  Both methods yielded consistent results. Example fits around H$\beta$ and the [O~III] doublet are shown in Figure~\ref{fig:sample_spec}. These line fluxes were then increased by 5\% to compensate for the fact that our grism extraction apertures (which were typically $2\arcsec$ to $4\arcsec$ in diameter) enclosed only 93\% to 97\% of the spectral flux for a point source.\footnote[9]{www.stsci.edu/hst/wfc3/documents/ISRs/WFC3-2011-05.pdf and www.stsci.edu/hst/wfc3/documents/ISRs/WFC3-2014-01.pdf}   The galaxies in our sample are not point sources but are small compared to the extraction aperture, with typical half-light radii of 0.25-0.50\arcsec\ \citep{hagen14}, indicating that our correction for extraction aperture flux loss is appropriate. Finally, these total H$\beta$ fluxes were converted to luminosity using the standard cosmology stated in the introduction. 

Note that we do not correct for underlying stellar absorption, which can affect determinations of the star formation rate \citep{moustakas06}.  
In the local universe, typical corrections for Balmer absorption are 
$\sim 4$~\AA\ in equivalent width (EW), but this number is a function of both the stellar population's 
age and IMF \citep{groves+12}.  Moreover, as seen in Figure~\ref{fig:ew_mag}, 4~\AA\  is relatively small 
compared to the measured equivalent widths of our objects.  Indeed, to verify that the effect is minor,
we repeated all our analyses while statistically adding 4~\AA~EW to each of our H$\beta$
measurements.  This has the effect of increasing all our H$\beta$ fluxes by an average of   
$\sim 10\%$, and increasing the significance of our findings.

\subsection{Gas-Phase Metallicity}

In addition to providing measurements of total luminosity, our Gaussian fits to [O~II], [Ne~III], H$\beta$, 
and [O~III] also allow the measurement of every system's gas phase metallicity, $12+ \log{\rm (O/H)}$.  The details of this analysis can be
found in \citet{gebhardt+14} including a catalog of the sources in this work, 
but in brief, we used the observed flux ratios and the polynomial relations in Table 4 of
\citet{maiolino+08} to estimate metallicity via the abundance sensitive diagnostics of [Ne~III] to [O~II], [O~III] 
to [O~II], and $R_{23}$ (([O~III] + [O~II]) / H$\beta$) \citep{zaritsky+94}. These estimates should be relatively reliable, as they mate
the $T_e$ ``direct'' methods, which are applicable to systems with $12 + \log({\rm O/H}) < 8.35$ to the photo-ionization models of \citet{kewley02}, which are 
useful for $12 + \log ({\rm O/H}) > 8.35$.  Some of these measures are 
relatively insensitive to extinction, while others are highly affected by reddening.  To account for this effect, we 
adopted a \citet{calzetti+00} extinction curve,  and computed gas-phase metallicity likelihood functions for fixed $E(B-V) = 0.2$.
The most likely system metallicity was adopted for our analysis.  Some abundance 
indicators, such as $R_{23}$, are double-valued, and many of our likelihood curves 
have two local maxima.  Fortunately, the use of the other diagnostics, such as [Ne~III] to [O~II] and [O~III] to 
[O~II], helped split this degeneracy, and usually led to the preference of one solution over the other.  To verify that our fixed
extinction value did not affect our results, we repeated our following analysis using all possible values of reddening
($0 < E(B-V) < 2$), and then marginalized over this uniform reddening distribution to derive the gas-phase metallicity likelihood
function.  The most likely system metallicities for all reddenings did not change our conclusions, only increased our metallicity error bars.
The distribution of our best-fit metallicities is shown in on the right side of Figure~\ref{fig:ew_mag}.

\begin{figure}[htp] 
\includegraphics[width=.45\textwidth]{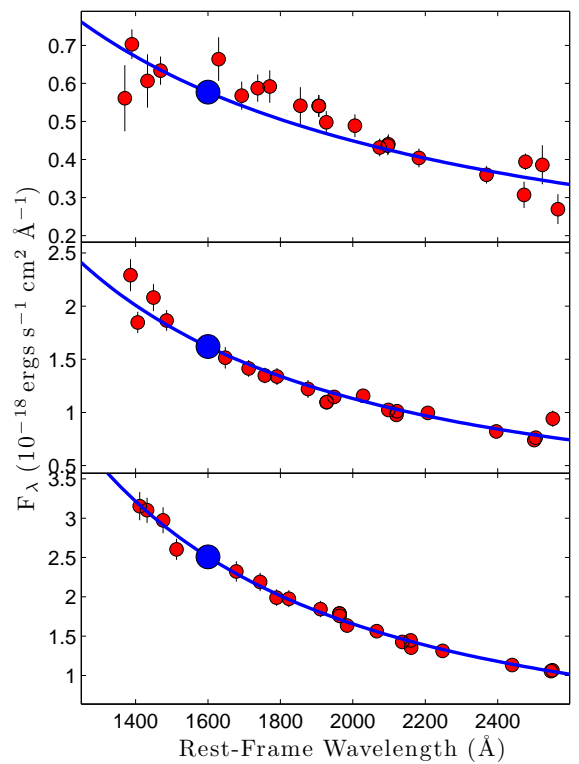}
\centering
\caption{Three examples of the photometric data available for the COSMOS and GOODS-S fields.  
From top to bottom, the three UV spectral energy distributions represent below average, average, and better than average
spectral energy distributions. The red points show the measured flux densities with their associated uncertainties, the blue curve displays
the best-fit power slope, and the large blue dot represents the best-fit flux density at 1600~\AA\null.  
Typically, for $z \sim 2$ galaxies, the rest-frame UV between 1250 and 2600~\AA\ is covered by $\gtrsim 15$ photometric bands; this allows us to accurately measure both the slope and normalization of the UV continuum.  }
\label{fig:goodss_beta}
\end{figure}

\subsection{UV Luminosity and Slope}

To obtain the rest-frame UV flux densities of our sources, we used the photometric catalogs produced 
by \citet{skelton2014}.   For star-forming populations, the wavelength range between 
$1250~{\rm\AA} < \lambda < 2600$~\AA\ samples the Rayleigh-Jeans portion of the hot stars' spectral energy
distributions.  Consequently, in the absence of reddening, the spectral slope across this
region should be relatively constant, i.e.,
\begin{equation}
F(\lambda) \propto \lambda^\beta
\end{equation}
where $\beta \sim -2.25$ for systems which have been forming stars at a constant rate for more than 
$\sim 10^8$~yr \citep{calzetti01}.  Values of $\beta$ larger than $-2.25$ can be attributed to the effects of internal
extinction, and, if the law of \citet{calzetti01} holds, $A_{1600} \sim 2.31 \, \Delta \beta$.

As Figure~\ref{fig:goodss_beta} illustrates, the photometry covering our program's fields is quite extensive \citep{skelton2014}.  
In the COSMOS field, broad- and intermediate-band measurements constitute a set of $\sim 19$ data points which can be fit
for $\beta$ and the observed flux density at 1600~\AA\ .  These bands include g, r, and i from CFHT \citep{erben09,hildebrandt09},
B$_J$, V$_J$, r+, i+ and 11 intermediate bands from Subaru \citep{taniguchi07,ilbert09}, and F606W from {\it HST}/ACS \citep{CANDELS,koekemoer11}.
In the GOODS-S field, there are $\sim 20$ data points covering the wavelength
range $1250~{\rm\AA} < \lambda_{\rm rest} < 2600$~\AA .  
These include B, V, R$_c$ and I from the WFI 2.2m \citep{erben05,hildebrandt06}, R from VLT/VIMOS \citep{Nonino09},
12 intermediate bands from Subaru \citep{cardamone10}, and F435W, F606W, and F775W from {\it HST}/ACS \citep{GOODS,CANDELS,koekemoer11}. 
Photometry in the GOODS-N field is not nearly so comprehensive, but it does include $\sim 9$ data points in the $z \sim 2$ rest-frame UV. 
These include G and R$_S$ from Keck/LRIS \citep{steidel03}, 
B$_J$, V$_J$, R, and i from Subaru \citep{capak04}, and F435W, F606W, and F775W from  {\it HST}/ACS \citep{GOODS,CANDELS,koekemoer11}.
Since the PSF-matched apertures of the \citet{skelton2014} photometric catalog are corrected for flux losses based on high-resolution imaging ({\it HST}/WFC3 F160W or F140W) at roughly the same wavelength as our grism observations, they serve as a good match to the total H$\beta$ fluxes provided by our grism measurements.

While complications may arise if the reddening curves contain a Milky-Way type bump at $\sim 2175$~\AA,
a careful examination of the COSMOS and GOODS-S photometry reveals no evidence for such a feature.  
This result is consistent with previous analyses, which have shown the bump to be less pronounced or 
non-existent in high equivalent-width objects such as those being studied \citep[\eg][]{kriek13}.  

\subsection{AGN Rejection}

Strong emission lines may be excited by the ionizing photons of hot stars, shocks in the ISM, and/or
active galactic nuclei (AGN).  In the local universe, diagnostic line ratios work quite well in
discriminating between these mechanisms \citep[\eg][]{bpt81, kewley+13}, but at $z \sim 2$,
key lines such as H$\alpha$ and [S~II] $\lambda\lambda 6716,6731$ shift out of the
range of the WFC3 grism.  Fortunately, there are medium and deep X-ray data over all three of 
our program fields \citep{elvis+09, alexander+03, xue11}.  At the redshifts considered here 
($1.90 < z < 2.35$), the X-rays associated with normal star formation are well below the limits of 
these surveys \citep{lehmer10}.  Consequently any emission-line galaxy whose position lies co-incident 
with an X-ray source is likely powered by an AGN\null.  

To identify the AGN, we therefore cross-correlated our $z \sim 2$ object catalog with the list of X-ray
sources found in the COSMOS, GOODS-N, and GOODS-S regions.   Only four of our emission-line 
galaxies lie within $2.5\arcsec$ of an X-ray source; these objects have been removed from our sample.  

\begin{figure}[htp] 
\includegraphics[width=.45\textwidth]{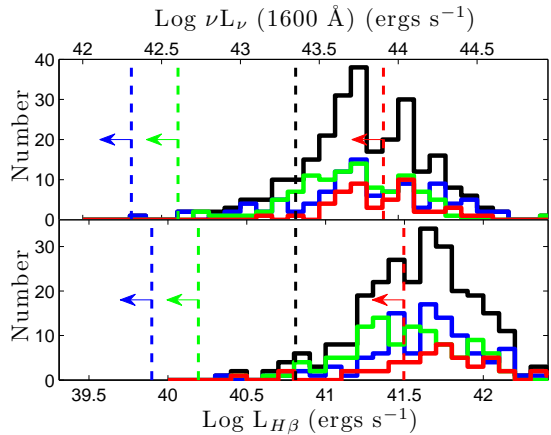}
\centering
\caption{The maximal AGN contribution to the observed H$\beta$ luminosity (bottom panel) and the
observed UV continuum luminosity (top panel).   The distributions for GOODS-S,
GOODS-N, COSMOS, and the combined dataset are displayed in blue, green, red, and
black, respectively.  The dashed vertical lines illustrate that maximum contribution from
AGN, and are based on the X-ray flux limits of the three survey fields
\citep{elvis+09, alexander+03, xue11}, an X-ray to optical power-law index of 1.9, and
a rest-frame UV power-law slope of $\alpha = -0.5$.   For
reference, the black dashed vertical line shows a SFR of $1 M_{\odot}$~yr$^{-1}$
\citep{hao+11, murphy+11}.  After excluding X-ray point sources from our analysis, we
find that any remaining AGN must have very little effect on the observed distributions of UV and H$\beta$ luminosities.}
\label{fig:agn}
\end{figure}

While we cannot exclude the possibility that low-luminosity AGN are contributing flux
to our survey, we can place limits on their effect.  To do this, we converted
the flux limits of the three X-ray surveys covering our fields \citep{elvis+09, alexander+03, xue11} into a 2 KeV luminosity density ($L_{\rm 2 KeV}$) using a 
power-law index of 1.9 and the online conversion tool, {\tt PIMMS}\footnote[10]{http://cxc.harvard.edu/toolkit/pimms.jsp}.  For 
AGN, there is a strong correlation between $L_{\rm 2 KeV}$ and the luminosity density at 2500\AA, $L_{2500}$.  We therefore used 
equation 6 in \citet{lusso10} to convert the $L_{\rm 2 KeV}$ limits into the maximum contribution ``normal'' AGN have to $L_{2500}$ .  We then assumed
a power-law slope of $\alpha = -0.5$ (i.e. L$_{\nu} \propto \nu^{\alpha}$; \citealp{vandenberk01}) to convert $L_{2500}$ to $L_{1600}$
as well as $L_{4861}$.  With an extra assumption that for AGN, the H$\beta$ equivalent width is typically $\sim$100 \AA\ \citep{binette93}, we
were able to estimate the maximal contribution AGN have on the observed UV and H$\beta$ luminosities. This can be seen in Figure~\ref{fig:agn}.
For GOODS-N and GOODS-S, AGN have very little effect on either luminosity measurement and can be neglected.  For COSMOS, the maximal
AGN contribution to the UV and H$\beta$ luminosities is less than or roughly equal to the median observed values, and the ratio of the contribution to $L_{{\rm H}\beta}$ to $L_{1600}$ is roughly what is expected from normal star-forming galaxies.  After excluding the individual X-ray sources from our analysis, it is clear that remaining AGN have very little effect on the observed UV and H$\beta$ luminosities and thus do not change our following SFR analysis. 

\vskip -0.4cm

%\vspace{-0.9cm}
\section{Star Formation Rates}

\subsection{Results}

%\begin{figure}[htp]
\begin{figure*}[htp]
\includegraphics[width=1\textwidth]{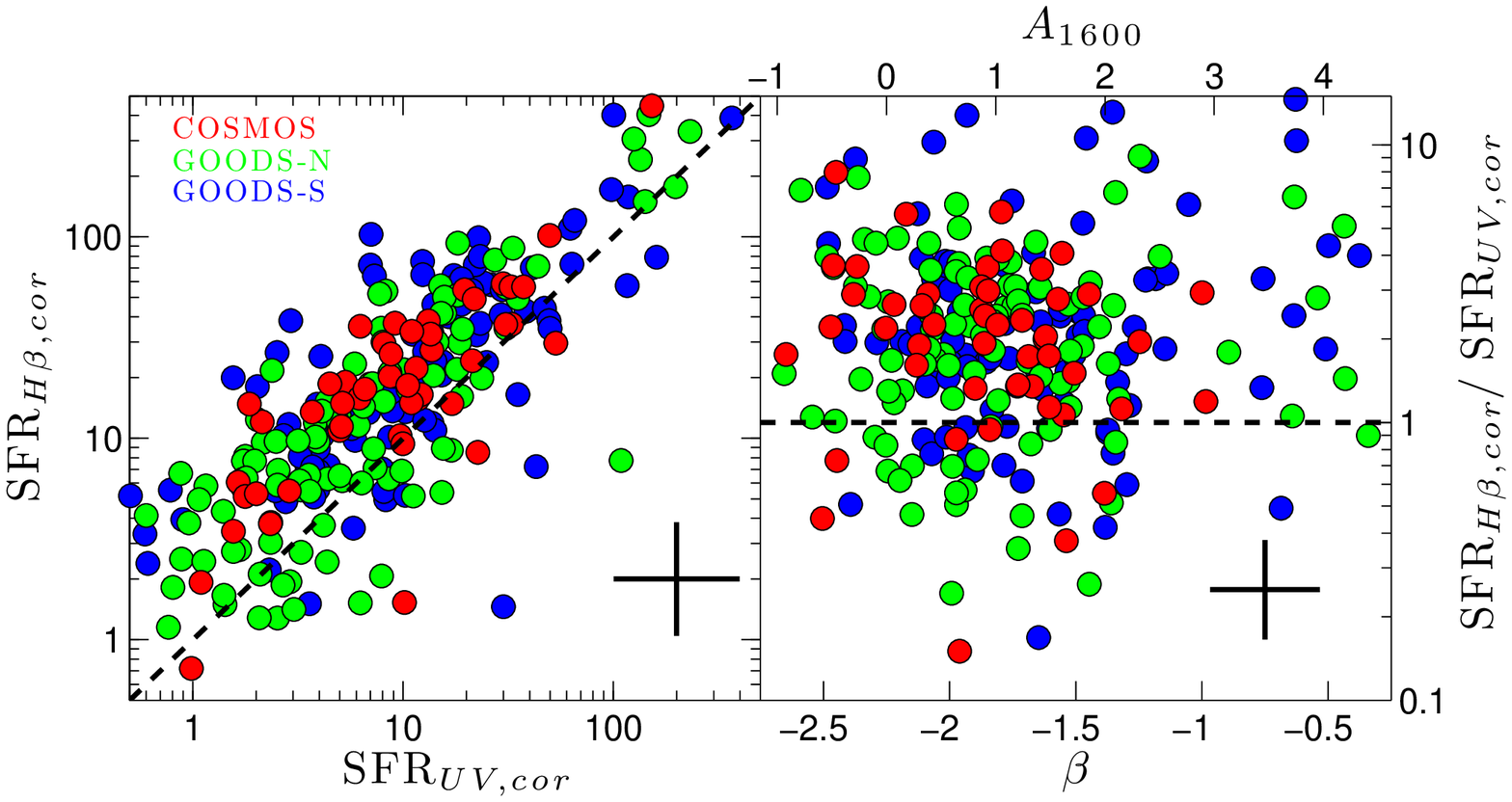}
\centering
\caption{A comparison of the H$\beta$ and UV star formation rates, using the SFR conversions of
\citet{hao+11} and \citet{murphy+11} (summarized in \citealp{kennicutt+12}) and a \citet{calzetti01} extinction law.   The left panel shows the extinction-corrected 
SFRs for COSMOS (red), GOODS-N (green), and GOODS-S (blue); the right panel displays the ratio
of these two measurements as a function of the UV continuum slope.  The typical 1-$\sigma$ uncertainties are shown as 
crosses.  On average, the H$\beta$ SFR is $\sim 1.8$ times that of the UV rate.  This result does
not depend on internal extinction.}
\label{fig:sfrplot}
%\end{figure}
\end{figure*}

Two of the most common tracers of star formation are the UV continuum and the hydrogen recombination lines
(\eg\ H$\beta$).  Both quantities are sensitive to the existence of short-lived, massive stars, but their systematics
are different.  The ultraviolet continuum at 1600~\AA\ measures the photospheric emission of stars with 
masses greater than a few $M_{\odot}$, hence the method records star formation over a time scale of 
$\sim 100$~Myr.   In contrast, the recombination lines of hydrogen are powered by photons shortward of 13.6~eV, with
\begin{equation}
L_{{\rm H}\beta} = Q({\rm H}^0) \left(\frac{{\alpha}^{\rm eff}_{\rm H\beta}}{{\alpha}_{B}}\right) 
h{\nu}_{{\rm H}\beta} \approx 4.81 \times 10^{-13}\ Q({\rm H}^0)\ {\rm ergs\ s^{-1}}~,
\label{eq:lhb}
\end{equation}
where ${\alpha}_{B}$ and ${\alpha}^{\rm eff}_{\rm H\beta}$ are the recombination coefficients for Case B and H$\beta$, respectively \citep{pengelly64, AGN3}. 
This number is virtually independent of temperature, density, and
metallicity; the luminosity of H$\beta$ only depends on the production rate of ionizing photons
($Q$), and the assumption that the interstellar medium is optically thick in the Lyman continuum.
Since the stars that produce these ionizing photons have higher masses ($M \gtrsim 15~M_{\odot}$) and 
shorter lifetimes ($t \lesssim 10$~Myr) than the stars traced by the rest-frame UV, the exact relationship
between the two star formation rate indicators can be complicated.  In particular, variables such as the 
IMF, the stellar metallicity, and the history of star formation can all effect
the observed ratio of the indicators.

The most common transformations between luminosity and star formation rate are those given by 
\citet{kennicutt98} and updated by \citet{kennicutt+12}.  These conversions, which were originally
tabulated in \citet{hao+11} and \citet{murphy+11}, are based on results from the STARBURST99 population
synthesis code \citep{SB99, vazquez05}, and assume a constant star formation history, solar metallicity, a \citet{kroupa01}
IMF, and a stellar population age of 10$^8$ years.  Using these relations, along with the assumption of 
Case~B recombination with an intrinsic H$\alpha$/H$\beta$ ratio of 2.86 \citep{brocklehurst71, AGN3},
a galaxy's star formation rate can be inferred from
\begin{equation}
\log~{\rm SFR}_{\rm H\beta} = \log L_{\rm H\beta} - 40.81~(M_{\odot}~{\rm yr}^{-1}),
\label{eq:shb}
\end{equation}
\begin{equation}
\log~{\rm SFR}_{\rm UV} = \log L_{UV} - 43.35~(M_{\odot}~{\rm yr}^{-1}),
\label{eq:suv}
\end{equation}
%(The FUV SFR relation found in \citet{kennicutt+12} is for $\sim1550$~\AA\ but is applicable 
%for 1600\AA\ to within 5\%.
where $L_{\rm H\beta}$ and $L_{UV}$ represent the total luminosities of H$\beta$ and the
UV continuum, after correcting for interstellar extinction.

This last issue can be problematic.   According to \citet{calzetti01}, in the local universe, the total
extinction (in magnitudes) at 1600~\AA\ is related to the slope of the UV continuum via
\begin{equation}
A_{1600} = \kappa_{\beta} \times (\beta - {\beta}_0),
\label{eq:ahb}
\end{equation}
where $\kappa_{\beta} = 2.31$, and $\beta_0 = -2.25$ for populations that have been forming stars 
for more than $\sim 10^8$~yr.  The connection between $\beta$ and the extinction of the H$\beta$
emission line is more tenuous, but again from \citet{calzetti01}
\begin{equation}
A_{\rm H\beta} = \zeta_{{\rm H}\beta} \, A_{1600}
\label{eq:auv}
\end{equation}
with $\zeta_{{\rm H}\beta} = 0.83$.  We begin by adopting these coefficients in our analysis, and
then test for variations by examining the systematics of the inferred star formation rates.

Figure~\ref{fig:sfrplot} compares the H$\beta$ and UV star formation rates for our sample of 260
$1.90 < z < 2.35$ galaxies in the COSMOS, GOODS-N, and GOODS-S regions.   From the figure, it
is clear that the relations summarized in \citet{kennicutt+12}, which typically produce a one-to-one correspondence in the 
$z \sim 0$ universe \citep[at least for SFR $\gtrsim 10^{-2} M_{\odot}$~yr$^{-1}$; see][]{lee+09}, do not work as 
well at $z \sim 2$.  On average, H$\beta$ star formation rates are $\sim 1.8$ times higher than those 
derived from the flux of the rest-frame UV\null.  Moreover, this effect cannot simply be attributed to 
extinction, as there is no correlation between the H$\beta$/UV SFR ratio and the value of $\beta$.  To identify the source of 
the discrepancy, we need to examine the effects that various assumptions have on the derived SFRs.

\subsection{Stellar Population Modeling}

%\begin{figure}[htp]
\begin{figure*}[htp]
\includegraphics[width=.95\textwidth]{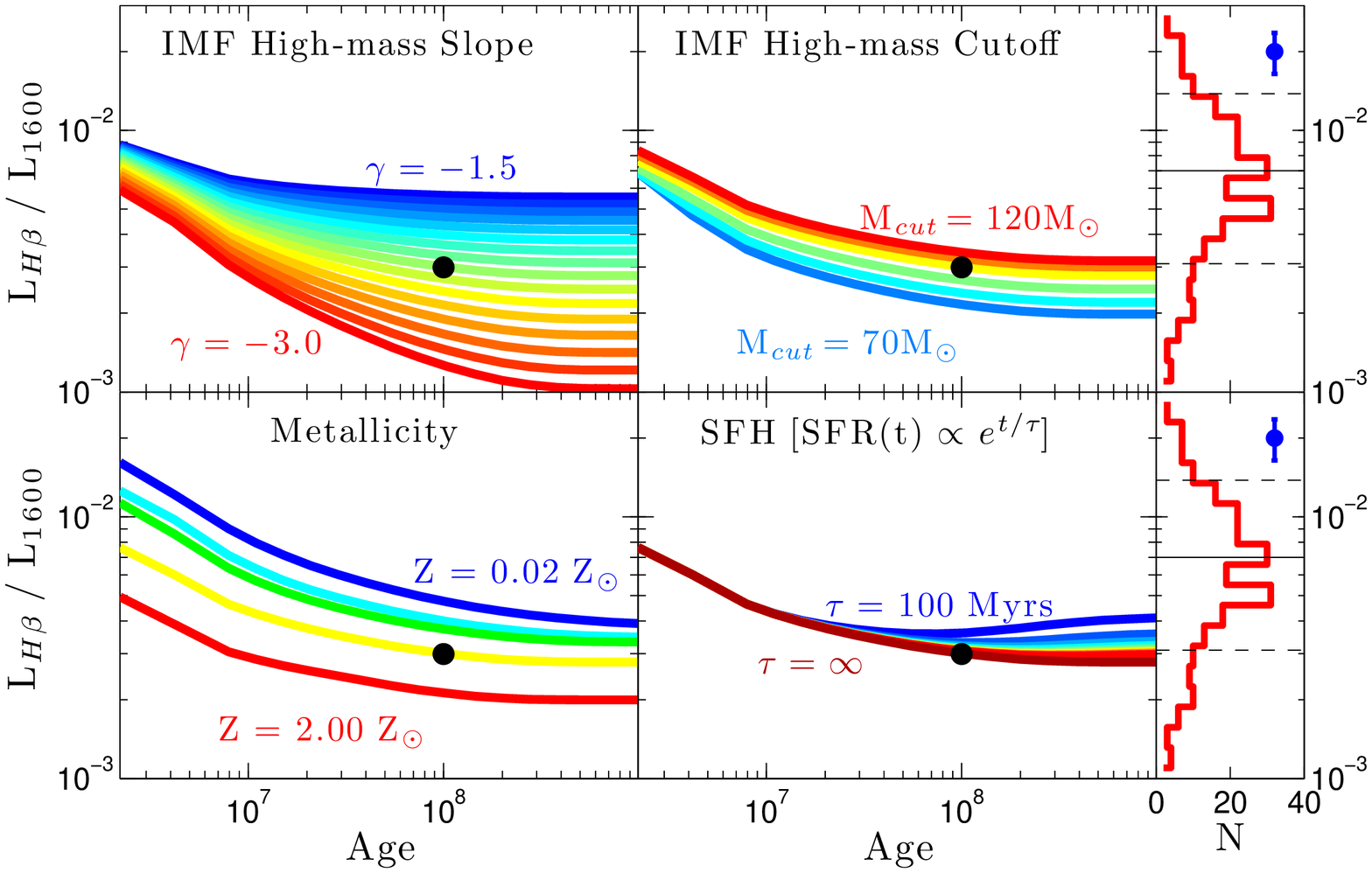}
\centering
\caption{The ratio of H$\beta$ luminosity to the luminosity density at 1600~\AA\ as a function of
age for a variety of STARBURST99 models.  The large black points represent the calibration model of \citet{hao+11} and \citet{murphy+11} (summarized in \citealp{kennicutt+12}), which assumes an IMF slope of $\gamma = -2.3$ \citep{salpeter55, kroupa01}, 
an upper mass cutoff of $100 M_{\odot}$, solar metallicity, and a constant star formation history. 
The various panels show the effect of changing the IMF (from $\gamma = -3.0$  to $-1.5$ in units of 0.1),
the high-mass cutoff (from $70 M_{\odot}$ to $120 M_{\odot}$ in units of $10 M_{\odot}$), the 
population metallicity (with $0.02 Z_{\odot}$, $0.20 Z_{\odot}$, $0.40 Z_{\odot}$, $1.00 Z_{\odot}$, and 
$2.00 Z_{\odot}$), and star formation history (with exponentially increasing $e$-folding time scales between 
$100~{\rm Myr} < \tau < 1$~Gyr in intervals of 100~Myr).  A constant SFR model ($\tau = \infty$) is also
presented.  The right-hand panels display the histogram of $L_{{\rm H}\beta} / L_{1600}$ for our sample of 
$z \sim 2$ star-forming galaxies, with the average 1-$\sigma$ error bar shown in blue.  The solid black 
horizontal line represents the median of the sample while the lower and upper dashed black lines are the
16$^{\rm th}$ and 
84$^{\rm th}$ percentiles.}
\label{fig:starburst_rat}
%\end{figure}
\end{figure*}

To model the systematics of the Balmer line and UV SFR indicators, we 
began with the assumptions that no ionizing photons are escaping into the intergalactic medium, 
and that dust in the H~II regions has, at most, a minor effect on the conversion of far-UV radiation to H$\beta$. 
The former assertion is probably quite good:  at $z \sim 3$, the escape fraction of Lyman
continuum photons is certainly less than 20\%, and probably below $\sim$ 5\% \citep{chen+07, iwata+09, 
vanzella+10, mostardi13}, and observations in the local universe suggest $f_{\rm esc} < 1\%$ 
\citep{adams+11}.   The latter assumption is also justifiable.  Enshrouding dust exists for only a short
period of time before O stars evaporate or evacuate it \citep{lada03}, and the dust that does survive likely
affects the UV and the Lyman continuum in roughly equal proportions.   In the metal-rich H~II regions
of the Milky Way, $\sim 25\%$ of far-UV photons may be absorbed by dust \citep[see Table 3 of][]{mckee97}, 
but in the metal-poor systems of the $z \sim 2$ universe, this fraction is likely to be much less.  We therefore
believe that this is not a large effect, and we can proceed to translate stellar emission into the 
observables of H$\beta$ and $L_{1600}$.

We examined the response of $L_{{\rm H}\beta}$ and $L_{\rm UV}$ to changes in stellar population by
first adopting as the baseline model of \citet{hao+11} and \citet{murphy+11} (summarized in \citealp{kennicutt+12}) SFR calibration, which uses solar metallicity,
a \citet{kroupa01} IMF, an upper mass cutoff of $100 M_{\odot}$, and 
a constant star formation history.  We then varied these parameters one at a time, using 
Version 6.0.4 of the
STARBURST99  \citep{SB99, vazquez05, leitherer10} population synthesis code with its Padova isochrones.  Figure~\ref{fig:starburst_rat} displays the results, which we discuss below.
This procedure is similar and consistent with previous works, albeit for H$\alpha$ to UV, which investigated the ratio produced by different IMFs (e.g.,~\citealp{meurer09}), stellar metallicities (e.g.,~\citealp{lee+09}), and star formation histories (e.g.,~\citealp{sullivan2000}).
  
\subsubsection{Initial Mass Function}

The IMF is usually assumed to be universal and power-law in form for masses
$M>M_{\odot}$.  As H$\beta$ and the UV are sensitive to different ranges of this function,
their predicted luminosity ratio will depend on the slope of the power-law ($\gamma$) and its upper
mass cutoff ($M_{\rm cut}$).  Varying these two parameters within a reasonable 
range can change the predicted $L_{{\rm H}\beta}-L_{1600}$ ratio by a factor of two or more.

%\begin{figure}[htp] 
\begin{figure*}[htp] 
\includegraphics[width=1\textwidth]{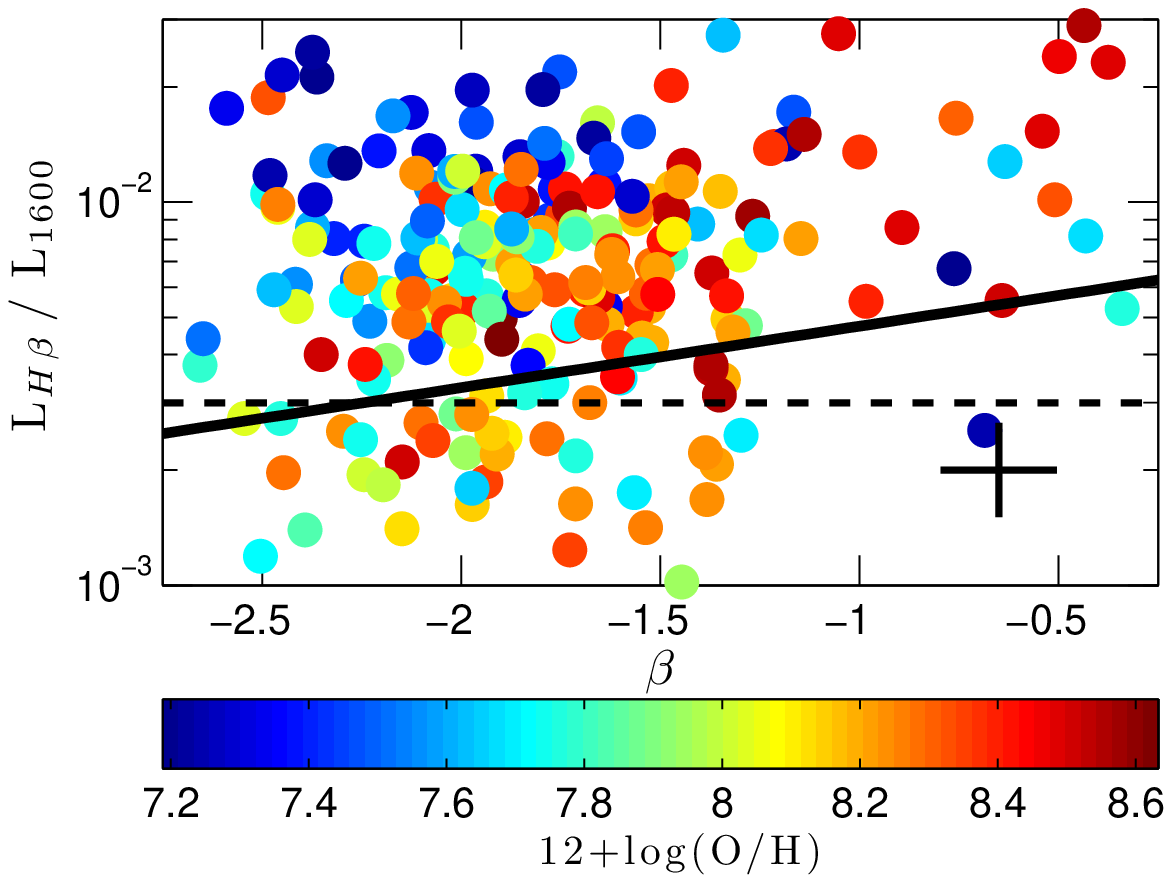}
\centering
\caption{The observed ratio of H$\beta$ luminosity to UV luminosity at 1600 \AA\ (before correcting
for reddening) as a function of continuum slope in the rest-frame UV\null.  The data points are colored by 
their best-fit gas-phase metallicity ($12 + \log {\rm O/H}$) and illustrate a correlation between metallicity and 
reddening, with the metal-rich systems being dustier.  A representative 1-$\sigma$ error bar is shown in the bottom right corner.
The dashed horizontal black line shows the value of
$L_{{\rm H}\beta} / L_{1600}$ expected from the SFR calibrations of \citet{hao+11} and \citet{murphy+11}, which are summarized in \citet{kennicutt+12}.   The solid black 
line couples these calibrations with the relation between $A_{1600}$ and H$\beta$ extinction detailed by
\citet{calzetti01}.    The anticipated anti-correlation between metallicity and $L_{{\rm H}\beta} / L_{\rm UV}$
is weak, but present.}
\label{fig:ratio_beta}
%\end{figure}
\end{figure*}

The top two panels of Figure~\ref{fig:starburst_rat} demonstrate this behavior.   As expected, a
flatter IMF implies relatively greater numbers of high-mass stars, and hence higher values for the 
$L_{{\rm H}\beta}/L_{1600}$.  In the Milky Way, $\gamma \sim -2.3$ \citep{salpeter55, kroupa01, chabrier03}, 
and this is the value used by \citet{hao+11} and \citet{murphy+11} (summarized in \citealp{kennicutt+12}) in their SFR calibrations.   However, despite recent advances
\citep[see][and references therein]{offner+13}, a firm theoretical understanding of the physics of the IMF
is still missing, thus its shape in $z \sim 2$ star-forming galaxies may be different.   Similarly, 
if the upper-mass limit to the main sequence is higher in our $z \sim 2$ systems, it will increase the luminosity 
of H$\beta$ more than that of the rest-frame UV continuum.

\subsubsection{Metallicity}

The metallicity of a stellar population affects the ratio of H$\beta$ to UV luminosity through the opacity and line blanketing 
in higher mass stars.  The lower the metallicity, the bluer the stellar 
population and the higher the ratio of $L_{{\rm H}\beta}$ to $L_{1600}$.  This effect can be seen in the 
bottom left panel of Figure~\ref{fig:starburst_rat}:  as we change the metallicity of the population from 0.02 
solar to twice solar, H$\beta$ becomes enhanced relative to the UV continuum.  As demonstrated by
\citet{gebhardt+14}, the gas-phase oxygen abundances for our sample of $z \sim 2$ galaxies range 
between $7.1 < 12+\log({\rm O/H}) <  8.7$ \citep[i.e., $0.025 \, Z_{\odot} < Z < Z_{\odot}$ with the solar
calibration of][]{asplund+09}, and the median value of the sample is $12+\log({\rm O/H}) = 8.06$ 
($Z \sim 0.2 Z_{\odot}$).   Thus, for star formation time scales larger than $\sim 100$~Myr, we should 
expect a $\sim 30\%$ increase in the $L_{{\rm H}\beta} / L_{1600}$ ratio compared to that given by \citet{hao+11} and \citet{murphy+11}.

Figure~\ref{fig:ratio_beta} displays the observed ratio of H$\beta$ to UV continuum luminosity as a function of
$\beta$, with the galaxies color-coded by their best-fit gas-phase metallicity.   An inspection of the figure
suggests the existence of a strong positive correlation between galactic extinction (as measured by the 
slope of the UV continuum) and oxygen abundance.  Indeed, a Spearman test confirms this trend,
as it rejects the null hypothesis that the two variables are uncorrelated with 99.9999\% confidence.   Of
course, a correlation between extinction and oxygen abundance makes sense, as the formation of 
dust should be tied to the presence of metals in the ISM (e.g.,~\citealp{garn10,reddy10}).  However, one would also expect a strong 
anti-correlation between metallicity and $L_{\rm H\beta} / L_{1600}$.  Since both the UV and H$\beta$ 
are powered by the energy emitted from young stars, and the $L_{\rm H\beta} / L_{1600}$ ratio is sensitive 
to metallicity, one would expect the two parameters to vary inversely with each other.  Indeed, the Spearman 
test rejects the null hypothesis with 99.8\% confidence. 

\subsubsection{Star Formation History}

\begin{figure}[htp] 
%\begin{figure*}[htp] 
\includegraphics[width=.45\textwidth]{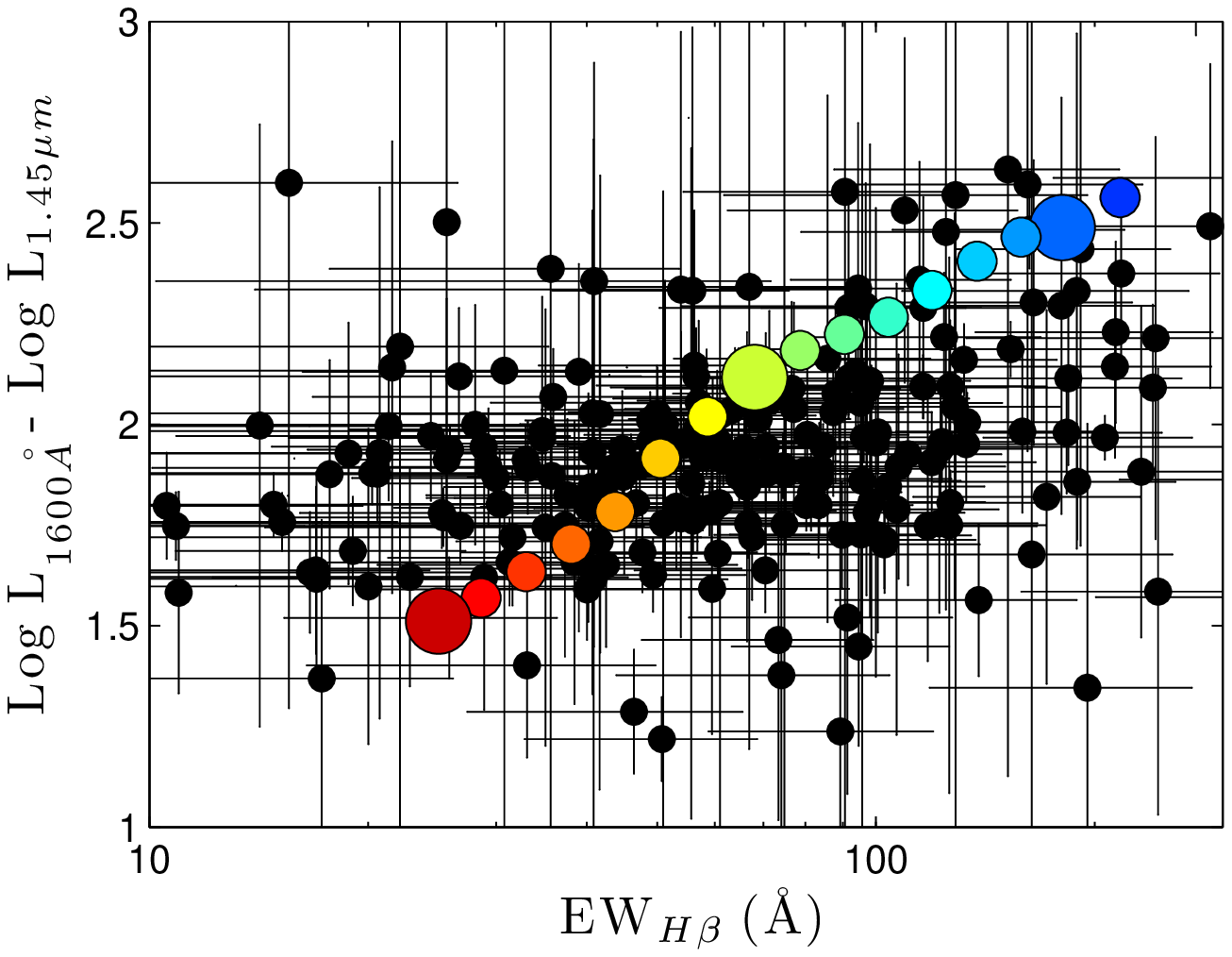}
\centering
\caption{The rest-frame equivalent width of H$\beta$ plotted against the rest-frame
UV minus IR color for our 3D-HST sample of galaxies.   Both quantities have been
de-reddened using the relations of \citet{calzetti01}; $k$-corrections have been
omitted, as these vary by less than 10\% over the redshift range of the survey.
Our 1-$\sigma$  error bars include measurement uncertainties and the propagation of
errors associated with extinction-correction.  The colored circles represent
STARBURST99 models with constant star formation, a Kroupa IMF,
$Z \sim 0.2 Z_{\odot}$, and ages from $7 \times 10^6$~yr (blue) to $10^9$~yr (red),
logarithmically spaced.  The larger circles in blue, yellow and red represent $10^7$,
$10^8$, and $10^9$~yr, respectively.   These constant star formation rate models
extend the range of our data both in color and equivalent width.}
\label{fig:sfh}
\end{figure}

Since H$\beta$ and the UV continuum are sensitive to different ranges of stellar mass, the age of the 
stellar population and the assumed star formation history are important for our
analysis.  In their calibration of star formation rate, \citet{hao+11} and \citet{murphy+11} assumed a constant star formation
rate history over a time scale greater than $10^8$~Myr.  To explore if this assumption is sufficient for our sample of galaxies, we examined
the relationship between two proxies of specific star formation rate (sSFR): EW$_{H\beta}$ and $\log L_{1600} - \log L_{1.45\mu m}$.  The former should be proportional to the mass-to-light ratio at 4861 \AA\ times the sSFR (see Equation 8 in \citealp{zeimann13}) while the latter weighs the newly formed population relative to the older, longer-lived stars.  As shown in Figure~\ref{fig:sfh}, a constant star formation history over a suite of ages from 7 $\times$ 10$^6$ years to 10$^9$ years cover the range of extinction-corrected, rest-frame EW$_{H\beta}$ as well as the extinction-corrected UV to IR color. 

In the literature, models with a bursty star formation history have been explored extensively in order to explain the scatter or offsets in Balmer line to UV SFR ratios (e.g. \citealp{sullivan2000,iglesias04,erb06,meurer09}).  The scatter in Figure~\ref{fig:sfrplot} however, can be explained simply by the combination of measurement error and the uncertainty in the extinction-correction. In other words, we need not invoke bursts to explain the diagram.  Moreover, the offset may be more simply explained by a range of ages for a constant star formation history.  The relation between H$\beta$ and UV SFR asymptotes for ages larger than 10$^8$ years, while for ages younger than that the H$\beta$ SFR will appear enhanced.  Our selection method may preferentially select young or newly-formed galaxies, some of which may have ages less than 10$^8$ years.  If we equate our extinction-corrected, rest-frame EW$_{H\beta}$ to an age using the models in Figure~\ref{fig:sfh}, we find that our sample selection biases the H$\beta$/UV SFR ratio by at most $\sim$10\%. 

Another consideration, as demonstrated by \citet{maraston+10}, is that
galaxies at $z \sim 2$ are better modeled with an increasing star formation rate history, and this has a
small effect on the predicted value of $L_{{\rm H}\beta}/L_{1600}$.  As illustrated in the bottom right panel of
Figure~\ref{fig:starburst_rat}, an exponentially increasing star formation rate with an $e$-folding timescale of 
$\tau = 300$~Myr will, over the course of a Gyr, increase $L_{{\rm H}\beta} / L_{1600}$ by $\sim 20\%$ over 
that of a constant star formation rate system.   Similarly, if star formation has only recently ignited, H$\beta$ will
again be boosted relative to the UV continuum.  Consequently, unless galaxies at $z \sim 2$ already have 
declining star formation rates, the H$\beta$ to UV ratio predicted by the \citet{kennicutt+12} calibration should
be a lower bound to the true value.

\subsection{Dust}

As nearly all stars form in clusters, the massive stars responsible for H$\beta$ and UV emission are initially 
co-located and enshrouded in high optical depth molecular clouds \citep{lada03}.  As the stellar population ages,
the most massive stars, which are primarily responsible for H$\beta$ emission, go supernovae and evacuate
much of their surrounding interstellar material, leaving the longer-lived B stars relatively unobscured.  
Consequently, as noted many times in the literature \citep[\eg][]{cha-fall00}, there can be a systematic 
difference between the extinction that affects H$\beta$ and that which reddens the UV continuum.  Moreover, this
offset can be a function of age, star formation history, galactic orientation, and dust composition.

\citet{cha-fall00} used a simple model of two separate environments, a birth cloud and a global ISM, to 
estimate the differential extinction seen by nebular emission and longer-lived stars.  Meanwhile,
\citet{calzetti+00} inferred an empirical relation between stellar and nebular extinction using
observations of eight nearby starburst systems.  Both studies reached the same conclusion:  in most
systems, $A_V$ for the gas should be roughly twice that of the stars.

For $z \sim 2$ systems, the wavelength coverage of 3D-HST survey does not extend out to H$\alpha$, and the wavelength separation between H$\beta$ and the higher-order Balmer lines is insufficient to obtain a robust
measure of extinction.  However, we can measure the amount of extinction affecting the stars via the slope of 
the rest-frame UV continuum.  STARBURST99 models confirm the results of \citet{calzetti01} that stellar 
populations dominated by a roughly steady-state number of young stars will have values of $\beta$ 
between $-2.4$ and $-2.2$, depending on the time scale for the on-going star formation.  Significantly, 
this number has little dependence on the IMF and stellar metallicity; any deviations from this
intrinsic slope must either arise from dust attenuation or, to a smaller extent, the SFR history of the stellar
population.  

\begin{figure*}[htp]
\includegraphics[width=1\textwidth]{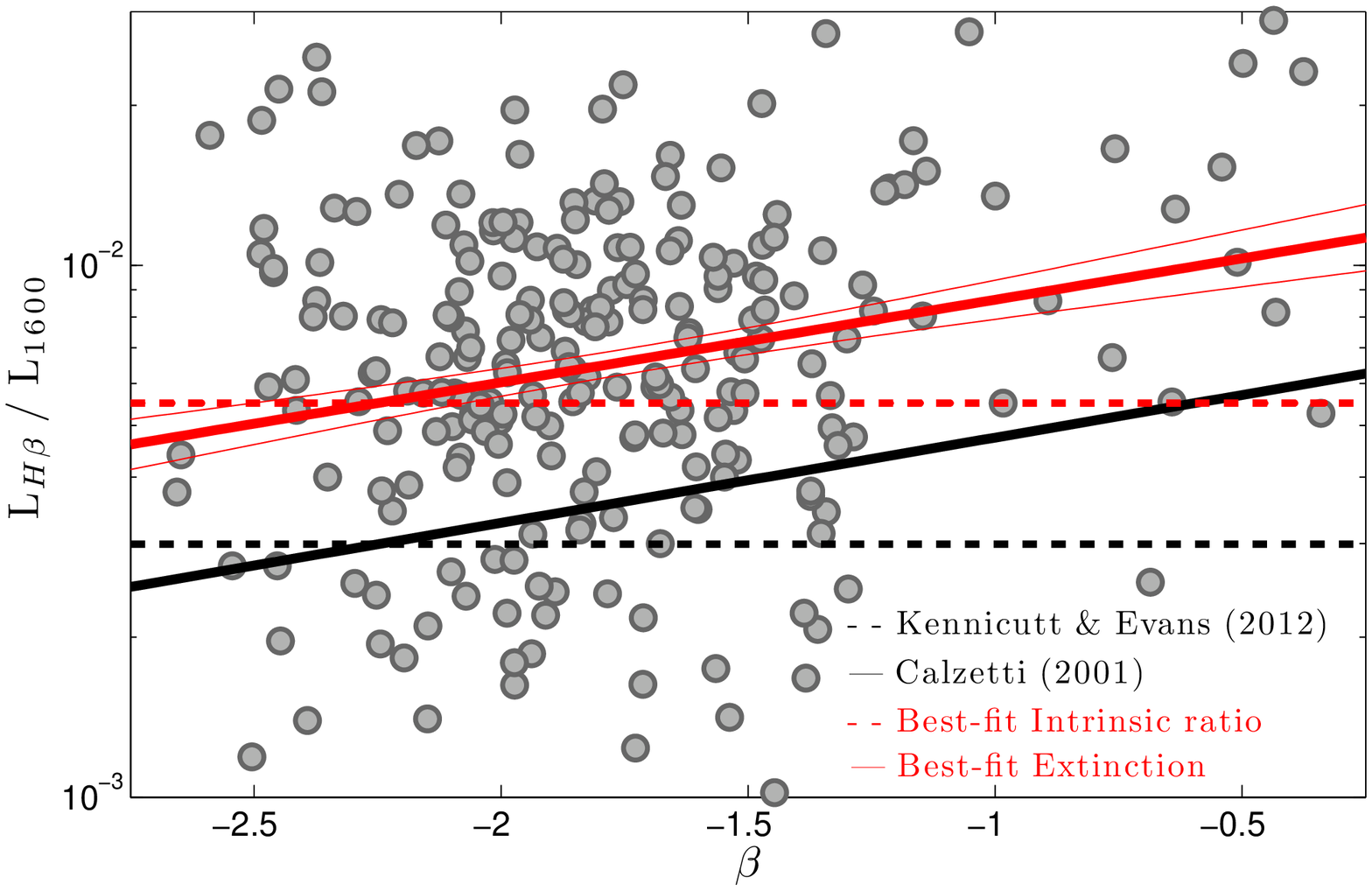}
\centering
\caption{As in Figure~\ref{fig:ratio_beta},  the observed ratio of H$\beta$ to UV luminosity (uncorrected for extinction) is plotted against the slope of the rest-frame UV continuum.  The black dashed line shows 
the expected ratio from \citet{hao+11} and \citet{murphy+11}, as summarized in \citet{kennicutt+12}, and the solid black line couples this number with the
\citet{calzetti01} extinction law so that the slope $m = 0.4 \times (1 - \zeta_{{\rm H}\beta})  \kappa_{\beta} =
0.162$.  The thick red line represents the best-fit linear regression, while the thinner red lines illustrate the
68\% confidence limits.  The intercept of this line with the zero-reddening value $\beta \sim -2.3$ demonstrates that $L_{{\rm H}\beta}/L_{\rm UV}$ is $1.84^{+0.17}_{-0.17}$ greater than that inferred from \citet{hao+11} and \citet{murphy+11}, thus excluding the local calibration with 99.9995\% confidence.  
This $z \sim 2$ zero-point is shown via the dashed red line.  Conversely, 
the slope of the best-fit line is $0.155 \pm 0.043$, which is consistent with \citet{calzetti01}.  For reference,
lower values for the slope indicate a larger differential extinction between gas and dust.}
\label{fig:dust_kenn}
%\end{figure}
\end{figure*}

In fact, the relationship between the observed ratio of $L_{H\beta} / L_{1600}$ and $\beta$ can reveal
more than just the extinction law.  The data and axes of Figure~\ref{fig:dust_kenn} are identical to those of 
Figure~\ref{fig:ratio_beta}, i.e., the figure plots the observed luminosities of H$\beta$ relative to the UV
continuum, uncorrected for extinction.   The best fit linear relation between this ratio and the slope of the
UV continuum (as determined by unweighted least squares) is shown in red.  The intercept of this line with 
$\beta \sim -2.3$ (i.e., where reddening should be minimal) provides information about the parameters of 
the underlying stellar population.  Conversely, the slope of the line, $m$, constrains the ratio of nebular to 
UV extinction through
\begin{equation}
\log {(L_{\rm H\beta} / L_{UV})} = \log {(L_{\rm H\beta} / L_{UV})_{\rm int}} + m (\beta - \beta_0) ,
\label{eq:slope_int}
\end{equation}
where,
\begin{equation}
m = 0.4 \times (1 - \zeta_{{\rm H}\beta}) \kappa_{\beta}~.
\label{eq:slope}
\end{equation}

As is illustrated in the figure, the intercept of the line with $\beta \sim -2.3$ is inconsistent with the SFR
calibrations of \citet{hao+11} and \citet{murphy+11}, as summarized in \citet{kennicutt+12}, at the 99.9995\% confidence level.  More specifically, 
$(L_{{\rm H}\beta}/L_{UV})_{\rm int}$ is $1.84^{+0.17}_{-0.17}$ larger than in the local universe, where 
the uncertainties represent 68\% confidence intervals.  Conversely, the slope of the relation, 
$m = 0.155 \pm 0.043$ is perfectly consistent with the value of 0.162 expected from a \citet{calzetti01} 
extinction law.  It therefore appears that at $z \sim 2$, $A_V$ for the gas is still approximately twice that of the 
stars. 
 
%\begin{figure}[htp]

%\begin{figure}[htp]
\begin{figure*}[htp]
\includegraphics[width=.95\textwidth]{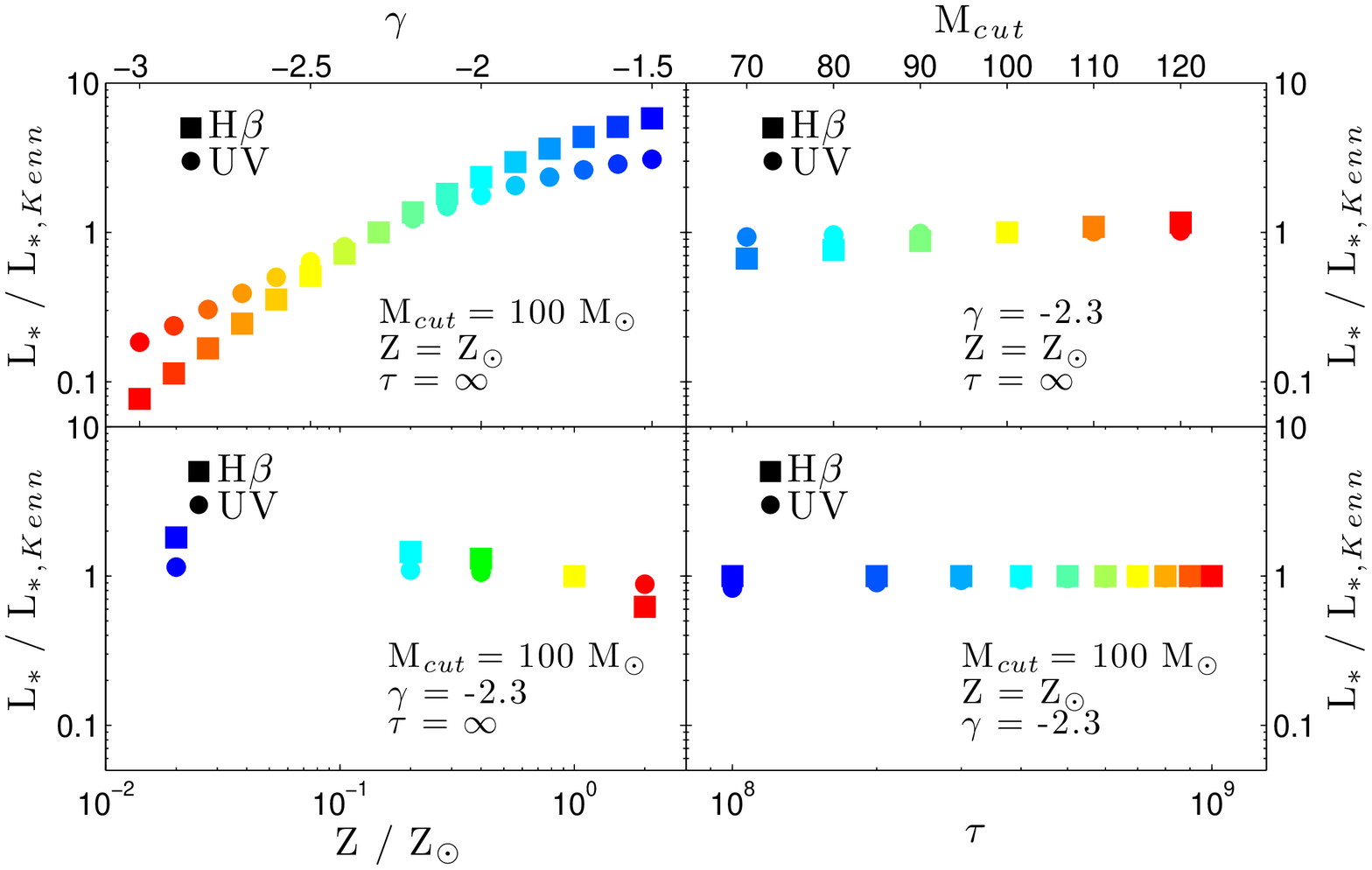}
\centering
\caption{The H$\beta$ (squares) and UV (circles) star formation rates compared to those predicted 
using the SFR calibrations of \citet{hao+11} and \citet{murphy+11}, as summarized in \citet{kennicutt+12}.  The colors are the same as Figure~\ref{fig:starburst_rat}, and the age of the stellar population is 10$^8$ years for all panels.  
Values of  $L_{*} / L_{*, {\rm Kenn}}$ greater than one imply that the \citet{kennicutt+12} relations will overestimate the true SFR, while values less than one indicate underestimates.  
As in Figure~\ref{fig:starburst_rat}, the top left panel varies the slope of the IMF, the top 
right panel shows the response to changes in the high-mass cutoff, the bottom left panel varies metallicity, 
and the bottom right panel changes the $e$-folding time scale for star formation.   These calculations were 
performed with the STARBURST99 population synthesis code, and do not include extinction.  In general, 
the strength of the hydrogen recombination lines is more sensitive to population parameters than the rest-frame UV.}
\label{fig:starburst_ind}
%\end{figure}
\end{figure*}

\section{Discussion} 

A wide range of methods are used for determining SFRs in different fields of astronomy.  Studies in the Local Group or the Milky 
Way galaxy may infer the star formation rate from the number of sources found in a molecular complex or 
H~II region via deep observations in the infrared or X-ray.  As each system may be in a different phase of
evolution, and have different external conditions and population parameters, the results of these studies can
be quite diverse \citep{chomiuk+11}.  In contrast, an extragalactic astronomer usually measures star formation 
over galaxy-wide scales, and must average over many of these differences.   Indeed, given the wide range of
properties observed for the star-forming complexes of the Milky Way \citep{feigelson+13} it is surprising the
degree of consistency that most SFR indicators exhibit, especially when the total SFR energy budget is 
well-tracked \citep{calzetti+07, kennicutt+12}.

In the local volume (out to 11~Mpc), UV and Balmer-line SFRs have been measured for a complete
set of galaxies extending all the way down to $M_B \lesssim -15$ \citep{lee+09}.  In these systems, 
measurements of the Balmer decrement and the total infrared luminosity have enabled independent determinations of
both the nebular and stellar reddening, thus allowing both SFR indicators to be tested in a 
variety of environments.  Interestingly, these data show that the SFR indicators summarized in \citet{kennicutt+12} do well
in systems with SFRs greater than $\sim 0.01 M_{\odot}$~yr$^{-1}$, but below this threshold, the relation
over-predicts the Balmer lines \citep{lee+09}.  Explanations for this offset include a variable IMF 
\citep[\eg][]{meurer09, boselli09,pflamm09}, stochasticity \citep{fumagalli11}, non-constant star formation histories
\citep{weisz12}, and leakage of ionizing photons into the intergalactic medium \citep{relano12}.   
Interestingly, this deficit for very low SFR systems is the exact opposite of what is seen at $z \sim 2$,
where H$\beta$ is enhanced relative to the UV continuum.  This suggest that very low SFRs in the local sample may be masking the effect metallicity
has on the ratio.  Unfortunately, as these SFRs are inaccessible for 3D-HST, a direct test of this hypothesis is not possible.
Still, it does indicate tension between observations and expectations. 

At high redshift, information from multiple indicators and constraints on extinction are limited.  
For this reason most high-$z$ studies simply assume the SFR calibrations of the local universe 
\citep[\eg][]{kennicutt98, kennicutt+12} and then allow the extinction law to float, as this is the most 
uncertain aspect of the analysis (e.g.,~\citealp{daddi07,forster09,wuyts13}).  The extinction that forces agreement between the Balmer line and 
UV SFRs is then adopted.   Not surprisingly, the results from such experiments at $1 < z < 3$ have varied.  
Some studies have found that extinction for the stars and gas are roughly equal, or, using the formalism
of this paper, $\zeta_{{\rm H}\beta} = 0.46$ \citep[\eg][]{erb06}.   Others analyses have concluded that the 
Balmer-line gas is typically extinguished roughly twice as much as the stars, and follows a \citet{calzetti01}
law with $E(B-V)_{\rm stars} = 0.44 \times E(B-V)_{\rm gas}$, or $\zeta_{{\rm H}\beta} = 1.05$
\citep[\eg][]{forster09, mannucci09, holden14}.  Still others suggest that the true relation is somewhere
in between \citep[\eg][]{wuyts13, price13}.  If we re-fit our data in Figure~\ref{fig:dust_kenn} while
restricting the $y$-intercept at $\beta = -2.25$ to match the SFR relations summarized in \citet{kennicutt+12} and use $\kappa_{\beta} = 1.99 - 2.31$ \citep{meurer99, calzetti01}, then we obtain $\zeta_{{\rm H}\beta} = 0.76 - 0.88$.
This is consistent with the intermediate case, where the gas is more extinguished than the stars but not 
twice as much.    
  
However, SFR calibrations must be treated with caution (e.g.,~\citealp{kennicutt83,kennicutt94,kennicutt09}).  Assumptions about the IMF, star formation history, 
and population metallicity all play a role in the transformation of observables  into estimates of star 
formation.   For example, Figure~\ref{fig:starburst_ind} uses the results of our STARBURST99 models to
demonstrate how H$\beta$ and the rest-frame UV continuum luminosity each respond to changes in the
commonly used assumptions that go into estimating SFRs.  The slope of the high-mass end of the initial mass
function has the greatest effect on our measurements.  Yet this parameter still lacks a theoretical 
understanding \citep[see][and references there within]{offner+13}, and is essentially unconstrained at $z \sim 2$.  More tractable is the response of the SFR to
changes in metallicity.   As shown in Figure~\ref{fig:starburst_ind}, the rest-frame UV is rather robust to shifts in the metal
abundance, as a population with $0.02 \, Z_{\odot}$ will only be $\sim 15$\% brighter than a 
corresponding solar-metallicity system. In contrast, the H$\beta$ luminosity of such a
metal-poor population will be larger by almost a factor of two, leading to a clear overestimate of the
star formation rate.  Previous works that have taken metallicity into account when calculating SFR (e.g.,~\citealp{lee02,brinchmann04,hunter04}) have found similar results.  Table~\ref{tab:sfr_conv} summarizes this fact by listing correction factors to the 
SFR relations summarized in \citet{kennicutt+12} as a function of metallicity.    For our 3D-HST sample, 
the gas-phase metallicities indicate a correction of $-0.16$ and $-0.04$~dex for our H$\beta$ and UV 
SFRs, respectively.  When this factor is applied to the data,  the expected $L_{{\rm H}\beta}$-$L_{\rm UV}$ 
ratio lies just outside of the 98\% confidence interval of Figure~\ref{fig:dust_kenn}.  Other corrections, such 
as that associated with an increasing star formation rate history, also make the expected intrinsic luminosity
more consistent with the data.

\begin{deluxetable}{c c c} 
%\tabletypesize{\scriptsize}
\tablecolumns{3}
\tablewidth{0pc}
\tablecaption{SFR Corrections}
\tablehead{\colhead{$Z / Z_{\odot}$}  & \colhead{Balmer emission} & \colhead{FUV}} 
\startdata
0.02 & $-$0.26 & $-$0.06  \\ 
0.20 & $-$0.16 & $-$0.04 \\ 
0.40 & $-$0.12 & $-$0.02 \\ 
1.00 & $+$0.00 & $+$0.00 \\ 
2.00 & $+$0.21 & $+$0.05  \\ 
\enddata
\tablecomments{
Additive logarithmic corrections to the SFR calibrations of \citet{hao+11}, \citet{murphy+11}, and \citet{kennicutt+12}.
}
\label{tab:sfr_conv} 
\end{deluxetable} 
\vskip +3cm
\section{Summary}

We use near-IR spectroscopy from 3D-HST and a wealth of photometric data in the GOODS-S, GOODS-N,
and COSMOS fields to compare the H$\beta$ and rest-frame UV extinction-corrected SFRs of 260 galaxies 
in the redshift range $1.90 < z < 2.35$.  Compared 
to the values expected from the UV luminosity density, our H$\beta$ SFRs are a factor of 
$\sim 1.8$ times higher than expected.  The lower metallicity of these $z \sim 2$ systems accounts for some 
of this offset, as models suggest that H$\beta$ should be enhanced by $\sim 45\%$, compared to only 
$\sim 10\%$ for the rest-frame UV\null.   Also, if star formation has only recently ignited, H$\beta$ will
 be elevated relative to the UV continuum.  Future IR spectroscopic surveys of $z > 2$ star forming systems 
will need to take these factors into account when interpreting their data.

Our observations also demonstrate that, as for the starburst galaxies of the local universe, the dust in
these $z \sim 2$ star forming systems extinguishes the stellar UV continuum more than optical
emission lines.  While the H$\beta$ and UV observations alone are insufficient to define the extinction law 
of these systems, we can determine a product which includes the total extinction at 1600~\AA\ 
and the ratio H$\beta$ to UV extinction.  Obviously, measurements of the Balmer decrement can break this
degeneracy, but even without these additional expensive observations, it is clear that our data
are in excellent agreement with a \citet{calzetti01} extinction law.   This result supports the premise that 
measurements of the rest-frame UV slope in $z \sim 2$ star-forming systems can be used to
estimate nebular reddening.

\acknowledgements
We would like to thank Eric Gawiser and Lucia Guaita for useful discussions during the preparation of this paper.   
We would also like to thank the anonymous referee who's careful reading and valuable comments greatly enhanced this study.
This work was supported via NSF through grant AST 09-26641 and AST 08-07873.  The Institute for Gravitation and 
the Cosmos is supported by the Eberly College of Science and the Office of the Senior Vice President for 
Research at the Pennsylvania State University.

\clearpage


\clearpage

\begin{thebibliography}{}

\bibitem[Adams \etal(2011)]{adams+11} Adams, J.J., Uson, J.M., Hill, G.J., \& MacQueen, P.J. 2011, 
\apj, 728, 107 

\bibitem[Alexander \etal(2003)]{alexander+03} Alexander, D.M., Bauer, F.E., Brandt, W.N., \etal\ 2003, 
\aj, 126, 539 

\bibitem[Asplund \etal(2009)]{asplund+09} Asplund, M., Grevesse, N., Sauval, A.J., \& Scott, P. 2009, 
\araa, 47, 481 

\bibitem[Baldwin \etal(1981)]{bpt81} Baldwin, J.A., Phillips, M.M., \& Terlevich, R. 1981, \pasp, 93, 5 

\bibitem[Bertin \& Arnouts(1996)]{bertin96} Bertin, E., \& Arnouts, S. 1996, \aaps, 117, 393 

\bibitem[Binette et al.(1993)]{binette93} Binette, L., Fosbury, R.~A., \& Parker, D.\ 1993, \pasp, 105, 1150 

%\bibitem[Boquien \etal(2012)]{boquien+12} Boquien, M., Buat, V., Boselli, A., \etal\  2012, \aap, 539, A145 

\bibitem[Boselli \etal(2009)]{boselli09} Boselli, A., Boissier, S., Cortese, L., \etal\  2009, \apj, 706, 1527 

\bibitem[Brammer \etal(2012)]{3DHST} Brammer, G.B., van Dokkum, P.G., Franx, M., \etal\  2012, \apjs, 200, 13

\bibitem[Brinchmann et al.(2004)]{brinchmann04} Brinchmann, J., 
Charlot, S., White, S.~D.~M., et al.\ 2004, \mnras, 351, 1151 

\bibitem[Brocklehurst(1971)]{brocklehurst71} Brocklehurst, M. 1971, \mnras, 153, 471 

\bibitem[Calzetti \etal(2000)]{calzetti+00} Calzetti, D., Armus, L., Bohlin, R.C., \etal\  2000, \apj, 533, 682 

\bibitem[Calzetti(2001)]{calzetti01} Calzetti, D.  2001, \pasp, 113, 1449 

\bibitem[Calzetti \etal(2007)]{calzetti+07} Calzetti, D., Kennicutt, R.C., Engelbracht, C.W., \etal\ 2007, 
\apj, 666, 870 

\bibitem[Capak et al.(2004)]{capak04} Capak, P., Cowie, L.~L., 
Hu, E.~M., et al.\ 2004, \aj, 127, 180 

%\bibitem[Capak \etal(2007)]{capak+07} Capak, P., Aussel, H., Ajiki, M., \etal\  2007, \apjs, 172, 99 

\bibitem[Cardamone \etal(2010)]{cardamone10} Cardamone, C.N., van Dokkum, P.G., Urry, C.M., \etal\
2010, \apjs, 189, 270 

%\bibitem[Cardelli \etal(1989)]{cardelli+89} Cardelli, J.A., Clayton, G.C., \& Mathis, J.S. 1989, \apj, 345, 245 

\bibitem[Chabrier(2003)]{chabrier03} Chabrier, G. 2003, \pasp, 115, 763 

\bibitem[Charlot \& Fall(2000)]{cha-fall00} Charlot, S., \& Fall, S.M. 2000, \apj, 539, 718 

\bibitem[Chen \etal(2007)]{chen+07} Chen, H.-W., Prochaska, J.X., \& Gnedin, N.Y. 2007, \apjl, 667, L125 

\bibitem[Chomiuk \& Povich(2011)]{chomiuk+11} Chomiuk, L., \& Povich, M.S. 2011, \aj, 142, 197

\bibitem[Daddi et al.(2007)]{daddi07} Daddi, E., Dickinson, M., 
Morrison, G., et al.\ 2007, \apj, 670, 156 

\bibitem[Dressel \etal(2014)]{dressel14} Dressel, L. 2014, Wide Field Camera 3 Instrument Handbook, Version 6.0 (Baltimore: STScI)

%\bibitem[Elbaz \etal(2011)]{elbaz11} Elbaz, D., Dickinson, M., Hwang, H.S., \etal\  2011, \aap, 533, A119

\bibitem[Elvis \etal(2009)]{elvis+09} Elvis, M., Civano, F., Vignali, C., \etal\  2009, \apjs, 184, 158 

\bibitem[Erb \etal(2006)]{erb06} Erb, D.K., Steidel, C.C., Shapley, A.E., \etal\  2006, \apj, 647, 128

\bibitem[Erben et al.(2005)]{erben05} Erben, T., Schirmer, M., 
Dietrich, J.~P., et al.\ 2005, Astronomische Nachrichten, 326, 432 

\bibitem[Erben et al.(2009)]{erben09} Erben, T., Hildebrandt, H., Lerchster, M., et al.\ 2009, \aap, 493, 1197 

\bibitem[Feigelson \etal(2013)]{feigelson+13} Feigelson, E.D., Townsley, L.K., Broos, P.S., \etal\  2013,
 \apjs, 209, 26 
 
\bibitem[Ford \etal(2003)]{ford03} Ford, H.C., Clampin, M., Hartig, G.F., \etal\  2003, \procspie, 4854, 81 
 
\bibitem[F{\"o}rster Schreiber \etal(2009)]{forster09} F{\"o}rster Schreiber, N.M., Genzel, R., Bouch{\'e}, N., 
\etal\  2009, \apj, 706, 1364 

\bibitem[Fruchter \etal(2009)]{fruchter09} Fruchter, A., Sosey, M., Hack, W., \etal\  2009, The
MultiDrizzle Handbook Version 3.0 (Baltimore, STScI)

\bibitem[Fumagalli \etal(2011)]{fumagalli11} Fumagalli, M., da Silva, R.L., \& Krumholz, M.R. 2011, \apjl, 
741, L26 

\bibitem[Garn 
\& Best(2010)]{garn10} Garn, T., \& Best, P.~N.\ 2010, \mnras, 409, 421 

\bibitem[Gebhardt \etal(2014)]{gebhardt+14} Gebhardt, H., Zeimann, G.R., Ciardullo, R., \& Gronwall, C.
2014, in preparation

\bibitem[Giavalisco \etal(2004)]{GOODS} Giavalisco, M., Ferguson, H.C., Koekemoer, A.M., \etal\ 
2004, \apjl, 600, L93 

\bibitem[Grogin \etal(2011)]{CANDELS} Grogin, N.A., Kocevski, D.D., Faber, S.M., \etal\  2011, 
\apjs, 197, 35 

\bibitem[Groves \etal(2012)]{groves+12} Groves, B., Brinchmann, J., \& Walcher, C.J. 2012, \mnras, 419, 1402 

\bibitem[Guo \etal(2013)]{guo+13} Guo, Y., Ferguson, H.C., Giavalisco, M., \etal\  2013, \apjs, 207, 24 

\bibitem[Hagen \etal(2014)]{hagen14} Hagen, A., Zeimann, G.R., Ciardullo, R., \& Gronwall, C.
2014, in preparation

\bibitem[Hao \etal(2011)]{hao+11} Hao, C.N., Kennicutt, R.C., Johnson, B.D., \etal\ 2011, \apj, 741, 124

\bibitem[Hildebrandt et 
al.(2006)]{hildebrandt06} Hildebrandt, H., Erben, T., Dietrich, J.~P., et al.\ 2006, \aap, 452, 1121 

\bibitem[Hildebrandt et 
al.(2009)]{hildebrandt09} Hildebrandt, H., Pielorz, J., Erben, T., et al.\ 2009, \aap, 498, 725 

\bibitem[Holden \etal(2014)]{holden14} Holden, B.P., Oesch, P.A., Gonzalez, V.G., \etal\ 2014, submitted to ApJ
(arXiv:1401.5490)

\bibitem[Hunter 
\& Elmegreen(2004)]{hunter04} Hunter, D.~A., \& Elmegreen, B.~G.\ 2004, \aj, 128, 2170

\bibitem[Iglesias-P{\'a}ramo \etal(2004)]{iglesias04} Iglesias-P{\'a}ramo, J., Boselli, A., Gavazzi, G., \& Zaccardo, A.\ 2004, \aap, 421, 887

\bibitem[Ilbert \etal(2009)]{ilbert09} Ilbert, O., Capak, P., Salvato, M., \etal\  2009, \apj, 690, 1236 

\bibitem[Iwata \etal(2009)]{iwata+09} Iwata, I., Inoue, A.K., Matsuda, Y., \etal\  2009, \apj, 692, 1287 

\bibitem[Kennicutt(1983)]{kennicutt83} Kennicutt, R.~C., Jr.\ 1983, 
\apj, 272, 54 

\bibitem[Kennicutt et al.(1994)]{kennicutt94} Kennicutt, R.~C., 
Jr., Tamblyn, P., \& Congdon, C.~E.\ 1994, \apj, 435, 22 

\bibitem[Kennicutt(1998)]{kennicutt98} Kennicutt, R.C., Jr. 1998, \araa, 36, 189 

\bibitem[Kennicutt \etal(2009)]{kennicutt09} Kennicutt, R.C., Jr., Hao, C.N., Calzetti, D., \etal\  2009, 
\apj, 703, 1672 

\bibitem[Kennicutt \& Evans(2012)]{kennicutt+12} Kennicutt, R.C., \& Evans, N.J. 2012, \araa, 50, 531

\bibitem[Kewley \& Dopita(2002)]{kewley02} Kewley, L.~J., \& Dopita, M.~A.\ 2002, \apjs, 142, 35 

\bibitem[Kewley \etal(2013)]{kewley+13} Kewley, L.J., Dopita, M.A., Leitherer, C., \etal\ 2013, \apj, 774, 100 

\bibitem[Koekemoer et al.(2011)]{koekemoer11} Koekemoer, A.~M., 
Faber, S.~M., Ferguson, H.~C., et al.\ 2011, \apjs, 197, 36 

\bibitem[Kriek \& Conroy(2013)]{kriek13} Kriek, M., \& Conroy, C. 2013, \apjl, 775, L16 

\bibitem[Kroupa(2001)]{kroupa01} Kroupa, P. 2001, \mnras, 322, 231

\bibitem[K{\"u}mmel \etal(2009)]{kummel09} K{\"u}mmel, M., Walsh, J.R., Pirzkal, N., Kuntschner, H., 
\& Pasquali, A. 2009, \pasp, 121, 59

\bibitem[Lada \& Lada(2003)]{lada03} Lada, C.J., \& Lada, E.A.  2003, \araa, 41, 57 

\bibitem[Lee et al.(2002)]{lee02} Lee, J.~C., Salzer, J.~J., 
Impey, C., Thuan, T.~X., \& Gronwall, C.\ 2002, \aj, 124, 3088 

\bibitem[Lee \etal(2009)]{lee+09} Lee, J.C., Gil de Paz, A., Tremonti, C., \etal\  2009, \apj, 706, 599 

%\bibitem[Lee \etal(2011)]{lee11} Lee, J.C., Gil de Paz, A., Kennicutt, R.C., Jr., \etal\  2011, \apjs, 192, 6 

\bibitem[Lehmer \etal(2010)]{lehmer10} Lehmer, B.D., Alexander, D.M., Bauer, F.E., \etal\  2010, \apj, 724, 559

\bibitem[Leitherer \etal(1999)]{SB99} Leitherer, C., Schaerer, D., Goldader, J.D., \etal\  1999, \apjs, 123, 3 

\bibitem[Leitherer \etal(2010)]{leitherer10} Leitherer, C., Ortiz Ot{\'a}lvaro, P.A., Bresolin, F., \etal\  2010, \apjs, 189, 309 

%\bibitem[Leroy \etal(2012)]{leroy12} Leroy, A.K., Bigiel, F., de Blok, W.J.G., \etal\  2012, \aj, 144, 3 

\bibitem[Lusso et al.(2010)]{lusso10} Lusso, E., Comastri, A., Vignali, C., et al.\ 2010, \aap, 512, A34 

\bibitem[Madau \etal(1996)]{madau96} Madau, P., Ferguson, H.C., Dickinson, M.E., \etal\ 1996, \mnras, 283, 1388 

\bibitem[Maiolino \etal(2008)]{maiolino+08} Maiolino, R., Nagao, T., Grazian, A., \etal\ 2008, \aap, 488, 463 

\bibitem[Mannucci \etal(2009)]{mannucci09} Mannucci, F., Cresci, G., Maiolino, R., \etal\  2009, \mnras, 
398, 1915 

\bibitem[Maraston \etal(2010)]{maraston+10} Maraston, C., Pforr, J., Renzini, A., \etal\  2010, \mnras, 407, 830 

\bibitem[McKee \& Williams(1997)]{mckee97} McKee, C.F., \& Williams, J.P. 1997, \apj, 476, 144 

\bibitem[Meurer \etal(1999)]{meurer99} Meurer, G.R., Heckman, T.M., \& Calzetti, D. 1999, \apj, 521, 64 

\bibitem[Meurer \etal(2009)]{meurer09} Meurer, G.R., Wong, O.I., Kim, J.H., \etal\  2009, \apj, 695, 765 

\bibitem[Mostardi \etal(2013)]{mostardi13} Mostardi, R.E., Shapley, A.E., Nestor, D.B., \etal\  2013, \apj, 779, 65 

\bibitem[Moustakas et al.(2006)]{moustakas06} Moustakas, J., Kennicutt, R.~C., Jr., \& Tremonti, C.~A.\ 2006, \apj, 642, 775 

\bibitem[Murphy \etal(2011)]{murphy+11} Murphy, E.J., Condon, J.J., Schinnerer, E., \etal\  2011, \apj, 737, 67 

\bibitem[Nonino et al.(2009)]{Nonino09} Nonino, M., Dickinson, 
M., Rosati, P., et al.\ 2009, \apjs, 183, 244 

\bibitem[Offner \etal(2013)]{offner+13} Offner, S.S.R., Clark, P.C., Hennebelle, P., \etal\  2013,  in
Protostars and Planets VI, eds.~H. Beuther, \etal (Tucson: University of Arizona Press), in press
(arXiv:1312.5326)

\bibitem[Oke \& Gunn(1983)]{oke83} Oke, J.B., \& Gunn, J.E. 1983, \apj, 266, 713 

\bibitem[Osterbrock \& Ferland(2006)]{AGN3} Osterbrock, D.E., \& Ferland, G.J. 2006, Astrophysics of
Gaseous Nebulae and Active Galactic Nuclei, 2nd.~ed.~by D.E. Osterbrock \& G.J. Ferland (Sausalito,
CA: University Science Books)

\bibitem[Pengelly(1964)]{pengelly64} Pengelly, R.M. 1964, \mnras, 127, 145 

\bibitem[Pflamm-Altenburg et al.(2009)]{pflamm09} 
Pflamm-Altenburg, J., Weidner, C., \& Kroupa, P.\ 2009, \mnras, 395, 394 

\bibitem[Price \etal(2013)]{price13} Price, S.H., Kriek, M., Brammer, G.B., \etal\ 2013, submitted to ApJ (arXiv:1310.4177)

\bibitem[Reddy et al.(2010)]{reddy10} Reddy, N.~A., Erb, D.~K., 
Pettini, M., Steidel, C.~C., \& Shapley, A.~E.\ 2010, \apj, 712, 1070 

\bibitem[Rela{\~n}o \etal(2012)]{relano12} Rela{\~n}o, M., Kennicutt, R.C., Jr., Eldridge, J.J., Lee, J.C., 
\& Verley, S.  2012, \mnras, 423, 2933 

\bibitem[Salpeter(1955)]{salpeter55} Salpeter, E.E. 1955, \apj, 121, 161 

\bibitem[Scoville \etal(2007)]{COSMOS} Scoville, N., Aussel, H., Brusa, M., \etal\  2007, \apjs, 172, 1 

\bibitem[Skelton et al.(2014)]{skelton2014} Skelton, R.~E., 
Whitaker, K.~E., Momcheva, I.~G., et al.\ 2014, (arXiv:1403.3689) 

%\bibitem[Sobral \etal(2013)]{sobral13} Sobral, D., Smail, I., Best, P.N., \etal\  2013, \mnras, 428, 1128 

%\bibitem[Steidel \etal(1999)]{steidel99} Steidel, C.C., Adelberger, K.L., Giavalisco, M., Dickinson, M., 
%\& Pettini, M. 1999, \apj, 519, 1 

\bibitem[Steidel et al.(2003)]{steidel03} Steidel, C.~C., 
Adelberger, K.~L., Shapley, A.~E., et al.\ 2003, \apj, 592, 728 

\bibitem[Sullivan et al.(2000)]{sullivan2000} Sullivan, M., Treyer, 
M.~A., Ellis, R.~S., et al.\ 2000, \mnras, 312, 442

\bibitem[Taniguchi et al.(2007)]{taniguchi07} Taniguchi, Y., 
Scoville, N., Murayama, T., et al.\ 2007, \apjs, 172, 9 

\bibitem[Tinsley(1980)]{tinsley80} Tinsley, B.M. 1980, \fcp, 5, 287 

\bibitem[Vanzella \etal(2010)]{vanzella+10} Vanzella, E., Giavalisco, M., Inoue, A.K., \etal\  2010, \apj, 725, 1011 

\bibitem[Vanden Berk et al.(2001)]{vandenberk01} Vanden Berk, D.~E., Richards, G.~T., Bauer, A., et al.\ 2001, \aj, 122, 549 

\bibitem[V{\'a}zquez \& Leitherer(2005)]{vazquez05} V{\'a}zquez, G.A., \& Leitherer, C.  2005, \apj, 621, 695 

\bibitem[Weisz \etal(2012)]{weisz12} Weisz, D.R., Johnson, B.D., Johnson, L.C., \etal 2012, \apj, 744, 44 

\bibitem[Wuyts \etal(2013)]{wuyts13} Wuyts, S., F{\"o}rster Schreiber, N.M., Nelson, E.J., \etal\ 2013, \apj, 779, 135 

\bibitem[Xue \etal(2011)]{xue11} Xue, Y.Q., Luo, B., Brandt, W.N., \etal\  2011, \apjs, 195, 10 

\bibitem[Zaritsky \etal(1994)]{zaritsky+94} Zaritsky, D., Kennicutt, R.C., Jr., \& Huchra, J.P. 1994, \apj, 420, 87 

\bibitem[Zeimann et al.(2013)]{zeimann13} Zeimann, G.~R., 
Stanford, S.~A., Brodwin, M., et al.\ 2013, \apj, 779, 137 

\end{thebibliography}
\end{document}